\documentclass{SciPost}

\hypersetup{
    colorlinks,
    linkcolor={red!50!black},
    citecolor={blue!50!black},
    urlcolor={blue!80!black}
}

\usepackage[bitstream-charter]{mathdesign}
\usepackage{tikz}
\usepackage{url}
\usepackage{hyperref}

\DeclareSymbolFont{usualmathcal}{OMS}{cmsy}{m}{n}
\DeclareSymbolFontAlphabet{\mathcal}{usualmathcal}

\fancypagestyle{SPstyle}{
\fancyhf{}
\lhead{\colorbox{scipostblue}{\bf \color{white} ~SciPost Physics }}
\rhead{{\bf \color{scipostdeepblue} ~Submission }}

\fancyfoot[C]{\textbf{\thepage}}
}

\begin{document}

\pagestyle{SPstyle}

\begin{center}{\Large \textbf{\color{scipostdeepblue}{
Electrostatics of Chern junctions in a magnetic field\\
}}}\end{center}

\begin{center}\textbf{
Robin Durand\textsuperscript{1,2$\star$},
Pascal Simon\textsuperscript{1} and
Ion Garate\textsuperscript{2 $\dagger$}
}\end{center}

\begin{center}
{\bf 1} Laboratoire de Physique des Solides, Universit\'e Paris-Saclay, CNRS, Orsay 91405, France
\\
{\bf 2} D\'epartement de physique, Institut quantique and Regroupement Qu\'eb\'ecois sur les Mat\'eriaux de Pointe,  Universit\'{e} de Sherbrooke, Sherbrooke, Qu\'{e}bec J1K 2R1, Canada
\\[\baselineskip]
$\star$ \href{mailto:email1}{\small robin.durand@universite-paris-saclay.fr} 
$\dagger$ \href{mailto:email2}{\small ion.garate@usherbrooke.ca}
\end{center}

\section*{\color{scipostdeepblue}{Abstract}}
\textbf{\boldmath{%
Electrostatic control is central to the operation of semiconductor
junctions made from conventional materials. 
Here, we theoretically investigate the electrostatic properties of lateral junctions between two-dimensional (2D)
Chern semiconductors in a perpendicular
magnetic field. The spectral asymmetry of a Chern semiconductor acts as a magnetic-field-induced
effective dopant density that is dependent on the Chern number. As a consequence, two regions with different Chern
numbers acquire different chemical potentials at charge neutrality, leading to a built-in potential of topological origin across the junction. Using this built-in
potential as a boundary condition, we solve the 2D Poisson equation to obtain the electric potential and charge density profiles, as well as the capacitance, of Chern semiconductor junctions. Our results 
inform the interplay between electronic topology and electrostatics in devices made from 2D materials. 
}}

\vspace{\baselineskip}

\noindent\textcolor{white!90!black}{%
\fbox{\parbox{0.975\linewidth}{%
\textcolor{white!40!black}{\begin{tabular}{lr}%
  \begin{minipage}{0.6\textwidth}%
    {\small Copyright attribution to authors. \newline
    This work is a submission to SciPost Physics. \newline
    License information to appear upon publication. \newline
    Publication information to appear upon publication.}
  \end{minipage} & \begin{minipage}{0.4\textwidth}
    {\small Received Date \newline Accepted Date \newline Published Date}%
  \end{minipage}
\end{tabular}}
}}
}



\vspace{10pt}
\noindent\rule{\textwidth}{1pt}
\tableofcontents
\noindent\rule{\textwidth}{1pt}
\vspace{10pt}


\section{Introduction}
\label{sec:intro}

Junctions made of conventional semiconductors, such as silicon, are fundamental building blocks of modern electronic and optoelectronic devices \cite{sze2021physics}. 
In a conventional $p$-$n$ junction, the contact between two differently doped regions produces charge redistribution, band bending, and a built-in electrostatic potential. These effects underlie rectification, carrier separation, and depletion capacitance, which are central to the operation of diodes, transistors, photodetectors, and solar cells.

In the past two decades, the discovery of topological materials has forged a new understanding of solids in terms of quantum mechanics and topology  \cite{anirban202315,bernevig2013topological,wieder2022topological}. In these materials, the electronic energy bands and wave functions are characterized by nonzero integers known as topological invariants. 
These mathematical invariants manifest physically by virtue of electronic states that are localized at sample boundaries and can carry electrical current with reduced dissipation.
As such, topological materials are being considered as candidates for microelectronic applications \cite{gilbert2021topological,gall2021materials,zhai2024rise}.

Chern insulators constitute a paradigmatic example of topological material. They are two-dimensional (2D), their bulk electronic bands are characterized by an integer topological invariant (the Chern number) and they host chiral edge states \cite{ren2016topological, chang2023colloquium}. 
Although the technological potential of Chern insulator device architectures has been highlighted \cite{gilbert2025chern}, little attention has been paid thus far to their electrostatic properties.
In particular, basic questions such as how the Chern number may affect the electric potential and charge density in junction geometries remain unexplored. 
Yet this type of question is timely, both from fundamental and applied perspective, given that Chern insulator junctions have recently been realized in magnetic topological materials \cite{yasuda2017quantized,rosen2017chiral, ovchinnikov2022topological} and in twisted van der Waals materials \cite{zhang2024manipulation, huber2026optical}. 
Depending on the system, the spatial Chern number profile has been engineered through a variation of the sample thickness or through an application of magnetic fields, electric gates or optical pulses.

\begin{figure}[h!]
    \centering
    \includegraphics[width=0.7\linewidth]{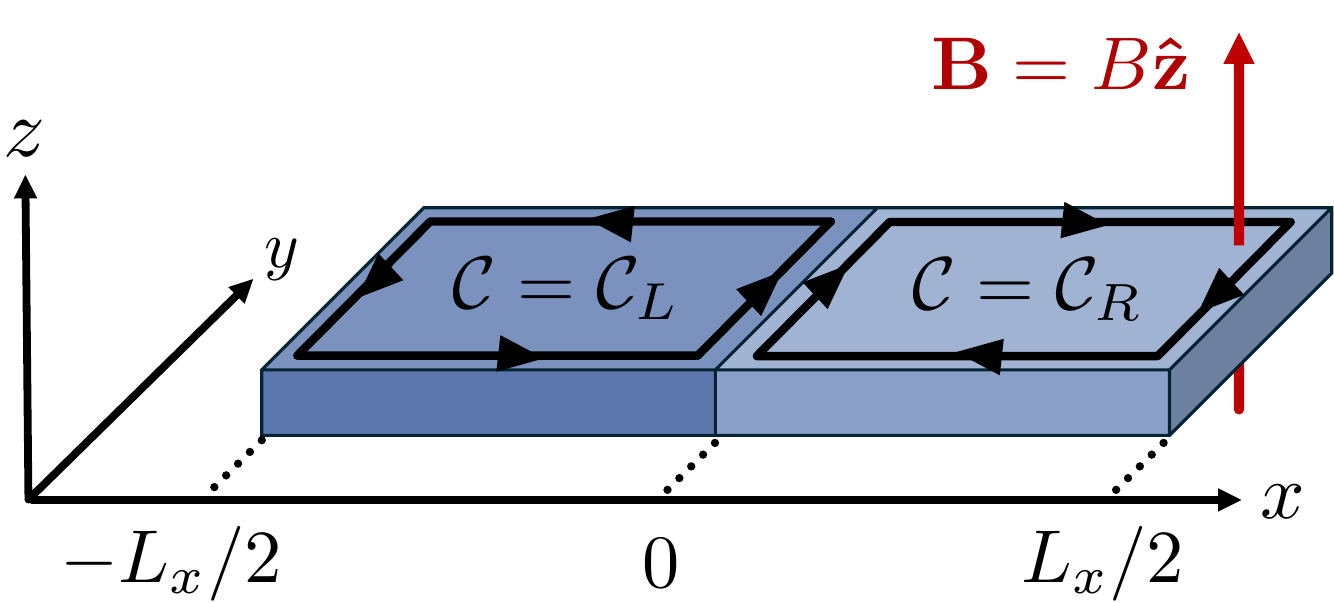}
    \caption{Schematic geometry of a Chern junction. The 2D system lies in the $xy$ plane and is divided into two semiconductor regions separated by an interface at $x=0$. For $x<0$, the material has a valence band Chern number $\mathcal{C}_L$, while for $x>0$ the valence band  Chern number is $\mathcal{C}_R$. The junction is translationally invariant along the $y$ direction.
    A uniform magnetic field $\mathbf{B}=B\hat{\mathbf z}$ is applied.
    The solid lines with arrows denote schematically the topological chiral edge states.}
    \label{fig:junction}
\end{figure}

It must be mentioned from the outset that the electrostatic properties of 2D lateral junctions are special even in the absence of topological invariants.
In a lateral 2D junction, the charge density is confined to an atomically thin layer, whereas the electric field extends into the surrounding three-dimensional (3D) space. In 2016, Yu {\em et al.} predicted that this nonlocal electrostatic environment leads to long-ranged charge redistribution and depletion widths with scaling laws that differ qualitatively from those of 3D junctions \cite{yu2016carrier}.
Those features were already remarked in a prescient work by Kane\cite{kane1988landau}.
Table \ref{tab:3D_v_2D} summarizes the main differences between the electrostatic properties of 3D and 2D junctions.
\begin{table}[]
    \centering
    \begin{tabular}{l|l|l}
    \hline\hline
         Properties & 3D junctions \cite{sze2021physics} & 2D junctions \cite{yu2016carrier, kane1988landau} \\
          \hline
         Charge density & 
         $\begin{array}{ll} \simeq-e c\,{\rm sgn}(x), & |x|<W \\\simeq0, & |x|>W\end{array}$  &  
         $\begin{array}{ll} \simeq-e c\,{\rm sgn}(x), & |x|<W \\ \simeq - e c  W/x, & |x|>W \end{array}$ \\
         Depletion width  & $W\propto\sqrt{|V_{\rm bi}/c|}$ & $W\propto |V_{\rm bi}/c|$\\
         Capacitance & $C_J\propto 1/W$  & $C_J\propto \ln(L_x/2 W)$\\
         \hline\hline
    \end{tabular}
    \caption{Comparison between the electrostatic properties of 3D and 2D $pn$  homojunctions made from nontopological materials, where $c$ ($-c$) is the net ionized dopant density at $x>0$ ($x<0$), $V_{\rm bi}$ is the built-in potential, $e$ is the proton charge, $W$ is the depletion width and $L_x$ is the system length.}
    \label{tab:3D_v_2D}
\end{table}
Although interesting, the results of Ref. \cite{yu2016carrier} apply to nontopological 2D materials. The objective of the present work is to find out how those results change for junctions made out of Chern insulators (Fig. \ref{fig:junction}), hereafter referred to as Chern junctions.

In short, it turns out that the Chern number influences the bulk electrostatic properties of a junction, provided that a perpendicular magnetic field $B$ is applied. 
Here, the role of the magnetic field is not to modify the electronic band topology, as we assume the value of the Chern number ${\cal C}$ to be independent of $B$. Instead, the role of the magnetic field is to render the bulk electrostatic properties of the junction dependent on ${\cal C}$. 

The need for a magnetic field can be anticipated from rudimentary symmetry arguments. Because the Chern number ${\cal C}$ is odd under time-reversal, it cannot directly impact the bulk electrostatic potential $\phi$, which is even under time-reversal.
However, the product ${\cal C} B$, which is even under time-reversal, is allowed to influence $\phi$ \footnote{One could argue that $|{\cal C}|$, being even under time-reversal, could affect $\phi$. In fact, that is what happens at the interfaces of a Chern insulator, where the 1D topological edge states generated by ${\cal C}\neq 0$ influence the local electrostatic potential and charge distribution. 
This influence is even under time-reversal, i.e. invariant under ${\cal C}\to-{\cal C}$, in the absence of an external magnetic field. In this work, we will be interested in elucidating bulk (2D, rather than 1D) electrostatic effects that are sensitive to the sign of the Chern number. For that purpose, a magnetic field will be necessary.}.
Likewise, a spatial gradient of ${\cal C} B$, which emerges naturally in a Chern junction, can influence the electric field\footnote{It is not necessary to have a junction with ${\cal C}_L\neq {\cal C}_R$ in order to create a spatial gradient in ${\cal C} B$. Indeed, one can consider a spatially homogeneous Chern insulator (${\cal C}_L= {\cal C}_R$) subject to a magnetic field that varies in space. Yet, this situation may be more difficult to realize in experiment.}. Quantifying this influence is the main purpose of the present work. 

We begin in Sec.~\ref{sec:model} by reviewing the minimal electronic structure of a Chern insulator in a square lattice, under a perpendicular magnetic field.
We then derive the expression for the charge density, which contains a topological contribution coming from the spectral asymmetry.
The spectral asymmetry shows up only in a magnetic field and qualitatively differentiates the electrostatic properties of ${\cal C}=0$ and ${\cal C}\neq 0$ systems.

In Sec. \ref{sec:poisson}, we discuss the Poisson equation that governs the electrostatics of Chern junctions and review the analytical solution provided by Ref. \cite{yu2016carrier} for materials with ${\cal C}=0$.
The key observation is that, from the electrostatic point of view, a Chern semiconductor behaves as if it carried a magnetic-field-induced carrier density. This density is not caused by dopant impurities, but by 
the spectral asymmetry.
Thus, when two regions with dissimilar Chern numbers are brought into contact, the Chern number difference forces the two regions to have different chemical potentials at charge neutrality and thereby generates a built-in potential of topological origin.
This built-in potential
is characterized theoretically in Sec. \ref{sec:built_in}.

In Sec. \ref{sec:charge_density_main}, the charge-density profile of the junction is derived.
The main finding is that the theoretical results of Ref. \cite{yu2016carrier} apply to Chern junctions as well, provided that the doping parameter $c$ and the built-in potential $V_{\rm bi}$ are appropriately modified to account for the Chern number, and provided that the
role of the topological edge states is considered in the proximity from the junction interface. 

In Sec. \ref{sec:capacitance}, the junction capacitance is extracted from the charge density profile.
This is an example of an experimentally measurable quantity that is sensitive to the electronic band topology.
The main text ends in Sec. \ref{sec:conclusions} with some conclusions, an outlook, and a brief qualitative comparison between the electrostatics of Chern junctions and the electrostatics of quantum Hall insulators with charge density gradients in 2D electron gases. In addition, the appendices contain derivations of certain results used in the main text.

\section{Chern semiconductor in a magnetic field}
\label{sec:model}

\subsection{Electronic structure}
\label{subsec:dirac_spectral}

A minimal two-band model for a Chern insulator at zero magnetic field is 
a half of the Bernevig-Hughes-Zhang (BHZ) 
Hamiltonian \cite{konig2008quantum}, namely
\begin{equation}
    H(\mathbf{k}) =
    \frac{A}{a}\bigl[\sin(k_x a)\,\sigma_x+\sin(k_y a)\,\sigma_y\bigr]
    + M_{\mathbf{k}}\,\sigma_z,
    \label{eq:hBHZ}
\end{equation}
where ${\bf k}=(k_x, k_y)$ is the wave vector in the $xy$ plane,  $a$ is the lattice constant for a  square lattice, $\sigma_\alpha$ are Pauli matrices in orbital space (the spin degree of freedom is neglected), $A$ is a constant 
and
\begin{equation}
    M_{\mathbf{k}}
    = M + \frac{2B_m}{a^2}\bigl[\cos(k_x a)+\cos(k_y a)-2\bigr],
    \label{eq:mass_k}
\end{equation}
with constant $M$ and $B_m$ \footnote{\label{note:real_space_H} The second quantized form of the tight-binding Hamiltonian in real space is
\begin{equation}
{\cal H} = -\frac{A}{2 a} \sum_{\bf R} \sum_{\boldsymbol{\delta}} \left( i \psi^\dagger_{\bf R} \boldsymbol{\sigma}\cdot\hat{\boldsymbol{\delta}}\psi_{{\bf R}+\boldsymbol{\delta}}+{\rm h.c.}\right)
+\frac{B_m}{a^2}\sum_{\bf R} \sum_{\boldsymbol{\delta}} \left(\psi^\dagger_{\bf R} \sigma_z \psi_{{\bf R}+\boldsymbol{\delta}}+{\rm h.c.}\right)+
\left(M-\frac{4 B_m}{a^2}\right)\sum_{\bf R} \psi^\dagger_{\bf R} \sigma_z \psi_{\bf R},
\end{equation}
where 
${\bf R}$ are the sites of the square lattice, $\boldsymbol{\delta}\in\{a \hat{\bf x}, a\hat{\bf y}\}$ are the nearest-neighbor vectors and $\psi^\dagger_{\bf R}$ and $\psi_{\bf R}$ are two-component creation and annihilation operators.}. 
For concreteness, we take $A>0$ and $B_m<0$.
For simplicity, we omit terms proportional to the identity matrix in Eq. (\ref{eq:hBHZ}), which can break particle-hole symmetry
In a junction made from two different Chern semiconductors, such terms may generate a finite built-in potential even in the absence of a magnetic field. 
But neglecting particle-hole symmetry-breaking terms in Eq. (\ref{eq:hBHZ}) does not qualitatively alter the main conclusions of the present work.

\begin{table}[h!]
    \centering
    \begin{tabular}{ c|c|c|c|c}
    \hline\hline
          & $\Gamma$ & $X$ & $Y$ & ${\cal M}$   \\
          \hline
         $M^{(i)}$ & $M$ & $M-4B_m/a^2$ & $M-4B_m/a^2$ & $M-8B_m/a^2$\\
         $v_x^{(i)}$ & $A/\hbar$  & $-A/\hbar$ & $A/\hbar$ & $-A/\hbar$\\
         $v_y^{(i)}$ & $A/\hbar$  & $A/\hbar$ & $-A/\hbar$ & $-A/\hbar$\\
         $\chi^{(i)}$ & +1 & -1 & -1 & +1\\ 
         \hline\hline
    \end{tabular}
    \caption{Dirac masses, velocities and chiralities at the high symmetry points of Eq. (\ref{eq:hBHZ}).}
    \label{tab:dirac}
\end{table}

From Eq. (\ref{eq:hBHZ}), massive Dirac fermions emerge at 
the four high-symmetry
points (or valleys) of the Brillouin zone: $\Gamma=(0,0)$, $X=(\pi/a,0)$, $Y=(0,\pi/a)$, ${\cal M}=(\pi/a,\pi/a)$.
Near these points, 
\begin{equation} 
\label{eq:hi}
H\simeq \hbar v_x^{(i)} q_x \sigma+\hbar v_y^{(i)} q_y \sigma_y+M^{(i)}\sigma_z+O(q_x^2, q_y^2),
\end{equation}
where ${\bf q}$ is the wave vector measured from the high-symmetry point $i$, $v_x^{(i)}$ and $v_y^{(i)}$ are the Dirac velocities and $M^{(i)}$ is the Dirac mass, listed in Table \ref{tab:dirac}.

\begin{figure}[h!]
    \centering
    \begin{tikzpicture}[
        x=1.00cm,
        y=1.05cm,
        >=latex,
        font=\small
    ]

        \def\xmin{-2}
        \def\xmax{10}
        \def\yaxis{-1.55}
        \def\r{2.1pt}

        \def\ChernLeft{-1}
        \def\ChernRight{+1}

        \colorlet{CzeroColor}{black}
        \colorlet{CplusColor}{blue!70!black}
        \colorlet{CminusColor}{red!75!black}

        \draw[-, thick]
            (\xmin-0.45,\yaxis) -- (\xmax+0.55,\yaxis)
            node[right] {\Large $\frac{Ma^2}{B_m}$};

        \draw[->, thick]
            (\xmin-0.45,\yaxis) -- (\xmin-0.45,1.85)
            node[above] {\Large $\mathcal{C}$};

        \foreach \y in {-1,0,1}{
            \draw[thick]
                (\xmin-0.53,\y) -- (\xmin-0.37,\y)
                node[left=2pt] {$\y$};
        }

        \foreach \x/\lab in {0/{0},4/{4},8/{8}}{
            \draw[thick]
                (\x,\yaxis-0.07) -- (\x,\yaxis+0.07)
                node[below=4pt] {$\lab$};
        }

        \foreach \x in {0,4,8}{
            \draw[dashed, gray]
                (\x,\yaxis+0.1) -- (\x,1.52);
        }

        \draw[very thick, CzeroColor]
            (\xmin,0) -- (-0.12,0);

        \draw[very thick, CminusColor]
            (0.12,\ChernLeft) -- (3.88,\ChernLeft);

        \draw[very thick, CplusColor]
            (4.12,\ChernRight) -- (7.88,\ChernRight);

        \draw[very thick, CzeroColor]
            (8.12,0) -- (\xmax,0);


        \draw[thick, CzeroColor]
            (0,0) ++(0,\r)
            arc[start angle=90,end angle=270,radius=\r];

        \draw[thick, CminusColor]
            (0,\ChernLeft) ++(0,\r)
            arc[start angle=90,end angle=-90,radius=\r];

        \draw[thick, CminusColor]
            (4,\ChernLeft) ++(0,\r)
            arc[start angle=90,end angle=270,radius=\r];

        \draw[thick, CplusColor]
            (4,\ChernRight) ++(0,\r)
            arc[start angle=90,end angle=-90,radius=\r];

        \draw[thick, CplusColor]
            (8,\ChernRight) ++(0,\r)
            arc[start angle=90,end angle=270,radius=\r];

        \draw[thick, CzeroColor]
            (8,0) ++(0,\r)
            arc[start angle=90,end angle=-90,radius=\r];

        \node[above=4pt, fill=white, inner sep=1.5pt, text=CzeroColor]
            at (-1.0,0)
            {$\mathcal{C}=0$};
        
        \node[above=4pt, fill=white, inner sep=1.5pt, text=CminusColor]
            at (2,\ChernLeft)
            {$\mathcal{C}=\ChernLeft$};
        
        \node[above=4pt, fill=white, inner sep=1.5pt, text=CplusColor]
            at (6,\ChernRight)
            {$\mathcal{C}=\ChernRight$};
        
        \node[above=4pt, fill=white, inner sep=1.5pt, text=CzeroColor]
            at (9.0,0)
            {$\mathcal{C}=0$};

        \node[align=center] at (0,1.73) {
            $\Gamma$
        };

        \node[align=center] at (4,1.73) {
            $X,\;Y$
        };

        \node[align=center] at (8,1.73) {
            $\mathcal{M}$
        };

    \end{tikzpicture}

    \caption{
     Topological phase diagram of the half-BHZ model for $B_m<0$, at zero magnetic field. The successive transitions are associated with gap closings at the high symmetry points indicated above the dotted lines. In particular, the transition between the $\mathcal{C}=+1$ and $\mathcal{C}=-1$ phases involves a simultaneous band inversion at two distinct points of the Brillouin zone, namely $X$ and $Y$, whereas the other transitions involve a gap closing at a single high-symmetry point.}
    \label{fig:half_bhz_phase_diagram}
\end{figure}
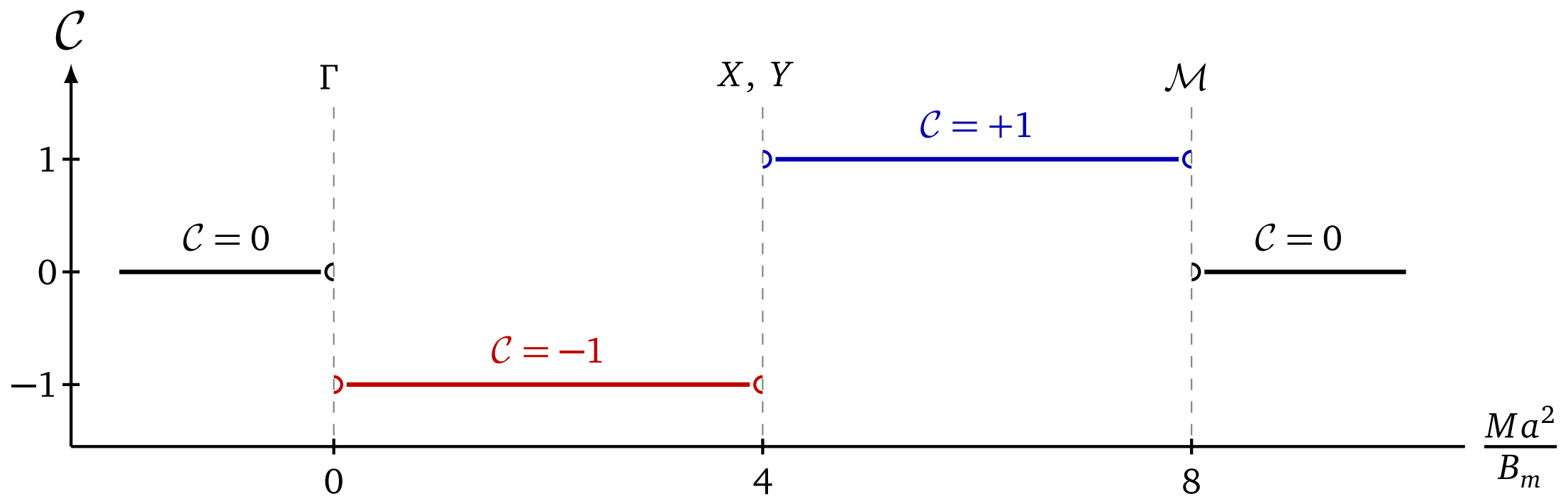

Each Dirac fermion of chirality 
\begin{equation}
\chi^{(i)}\equiv{\rm sgn}(v_x^{(i)} v_y^{(i)})
\end{equation} 
contributes a partial Chern number\cite{bernevig2013topological} 
\begin{equation}
    {\cal C}^{(i)}=\frac{1}{2}\chi^{(i)} {\rm sgn}(M^{(i)}).
    \end{equation}
    Summing them all, the Chern number of the valence band of Eq. (\ref{eq:hBHZ}) reads
\begin{equation}
    {\cal C} = \sum_i  {\cal C}^{(i)}= \frac{1}{2}\Bigl[
        \mathrm{sgn}(M)
        -2\,\mathrm{sgn}(M-4B_m)
        +\mathrm{sgn}(M-8B_m)
    \Bigr],
    \label{eq:chern}
\end{equation}
as illustrated in Fig. \ref{fig:half_bhz_phase_diagram}.

\begin{figure}[h!]
    \centering
    \includegraphics[width=\linewidth]{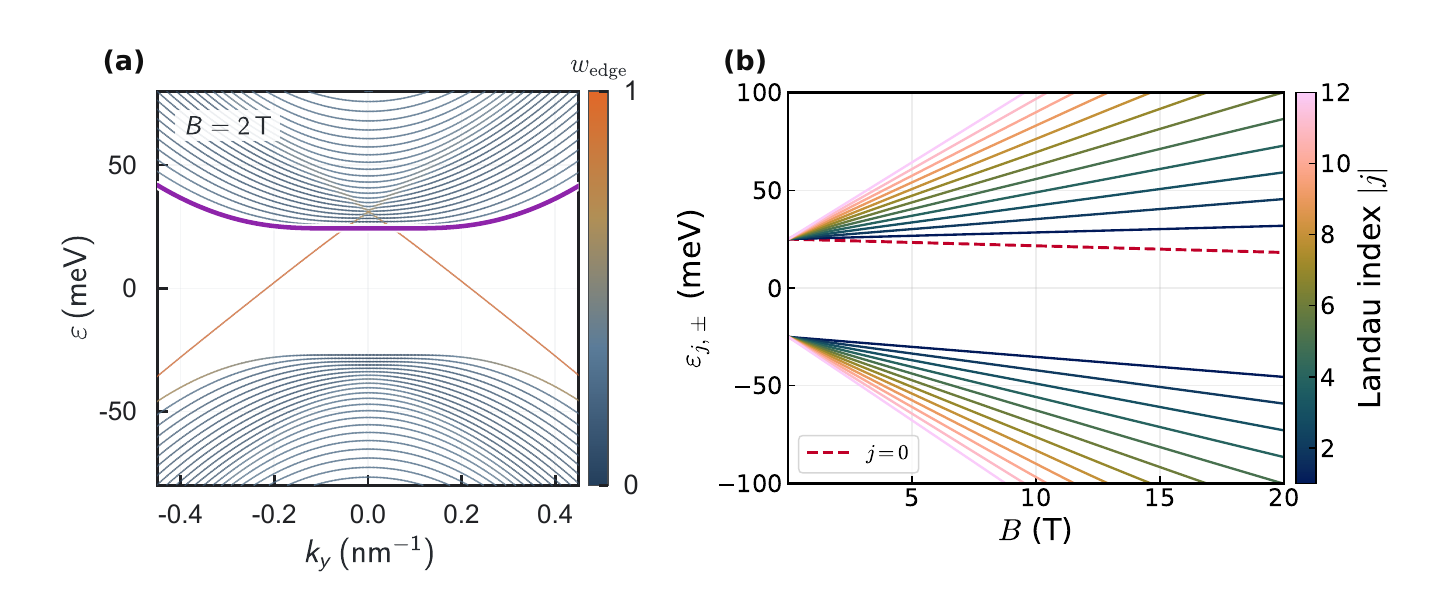}
\caption{
Landau-level spectrum of the half-BHZ Hamiltonian in a perpendicular magnetic field for the choice of band parameters $A=150\,\mathrm{meV.nm}$, $B_m=-225\,\mathrm{meV.nm^2}$, $M=-25\,\mathrm{meV}$, $a=1\,\mathrm{nm}$.
(a) Energy dispersion $\varepsilon(k_y)$ computed from the tight-binding model in a ribbon geometry, open along $x$ ($L_x=150\,\mathrm{nm}$) and periodic along $y$, at $B=2\,\mathrm{T}$. The color encodes the edge-localization weight $w_{\mathrm{edge}}=\sum_{x\in\mathrm{edges},\,\alpha}|\psi_{\alpha}(x)|^2$ where the edge region is defined as the outermost $20\times a$ sites $(20\,\mathrm{nm})$ on each side of the ribbon.
Orange branches crossing the bulk gap correspond to chiral edge states, while blue nearly flat bands correspond to bulk Landau levels. The purple line highlights the bulk anomalous $j=0$ Landau level, which lies in the conduction band for the present choice of band parameters.
(b) Landau fan diagram in the vicinity of the $\Gamma$ point, obtained from the continuum Hamiltonian including the quadratic correction proportional to $B_m$ (cf. App. \ref{app:LL_Bm}). The dashed red line denotes the anomalous $j=0$ Landau level, while the solid branches correspond to the $j>0$  Landau levels; the color encodes the Landau index $|j|$. The magnetic field strengths considered throughout the present work are weaker than those required to move an anomalous Landau level from the conduction band to the valence band (or vice versa). 
}
    \label{fig:dispersion}
\end{figure}

Next, a uniform perpendicular magnetic field $B$ is incorporated through ${\bf k}\to {\bf k} + e {\bf A}/\hbar$ \cite{bottcher2019survival, bottcher2020fate,bottcher2021fate},
with $-e<0$ the electron's charge and ${\bf A}=B x\hat{\bf y}$ the vector potential
\footnote{Note that the sign choices in Eq. (\ref{eq:hBHZ}) are not identical to those of Refs. \cite{bottcher2019survival, bottcher2020fate,bottcher2021fate}. In the actual numerical calculation, the Peierls substitution is implemented in the real space Hamiltonian of footnote \ref{note:real_space_H}, via
\begin{equation}
c^\dagger_{\bf R} \sigma_\alpha c_{{\bf R}+\boldsymbol{\delta}} \to e^{\frac{ i e}{\hbar} \int_{\bf R}^{{\bf R}+\boldsymbol{\delta}}{\bf A}\cdot d{\bf r}} c^\dagger_{\bf R} \sigma_\alpha c_{{\bf R}+\boldsymbol{\delta}}.    
\end{equation}}, such that ${\bf B} = B \hat{\bf z}$. 
Then, the electronic energy spectrum splits into Landau levels (Fig. \ref{fig:dispersion}(a)).

The low-energy part of the bulk energy spectrum can be modelled analytically in the Dirac fermion approximation (see e.g. the review article of M. O. Goerbig at \cite{miniatura2011ultracold}, also available at \cite{goerbig2009quantum}). 
Specifically, close to a high-symmetry point $i$ with chirality
$\chi^{(i)}$ and mass $M^{(i)}$, the ordinary Landau levels with integers $j\geq 1$ are
\begin{equation}
    \varepsilon_{j\geq 1,\pm}^{(i)}
    \simeq  \pm\sqrt{(M^{(i)})^2+\frac{2jA^2}{\ell_B^2}},
    \label{eq:LL_ordinary_dirac}
\end{equation}
where $\ell_B=\sqrt{\hbar/e |B|}$ is the magnetic length and the positive (negative) sign labels conduction-band (valence-band) Landau levels. 

A crucial feature of the bulk electronic dispersion is the
anomalous $j=0$ Landau level. Its energy equals 
\begin{equation}
\label{eq:LL_anom_dirac}
   \varepsilon_{0}^{(i)} \simeq -\chi^{(i)} M^{(i)} {\rm sgn}(B). 
\end{equation}
The anomalous level belongs to the conduction band when $C^{(i)} B<0$ and to the valence band when $C^{(i)}B>0$.
This unpaired level breaks the electron-hole symmetry of the Landau
level spectrum.

The half-BHZ curvature parameter $B_m$ produces
additional corrections to Eqs. (\ref{eq:LL_ordinary_dirac}) and (\ref{eq:LL_anom_dirac}) \cite{bottcher2021fate}, which are omitted here for brevity but are listed in Appendix~\ref{app:LL_Bm} and will be kept when needed
in the developments below.

In this work, we assume that the chemical potential is within the bulk gap when the external magnetic field is turned off, with thermally excited carriers in the bulk conduction and valence bands, much like in conventional semiconductors. 
For a weak enough magnetic field, the chemical potential remains well within the bulk gap. For a stronger field, the chemical potential approaches a band edge. These statements will be made more precise in Sec. \ref{subsec:chemical_potential_neutrality}.
In sum, for our purposes, the Chern insulator acts like a Chern semiconductor.

Aside from the bulk states described above, Fig. \ref{fig:dispersion} displays one-dimensional chiral edge states that cross the bulk energy gap. 
Though topological, these spatially localized (one-dimensional) states do not play a leading role in the bulk (2D) electrostatic properties of the junction, as will be shown in Sec. \ref{subsec:interface_states}. What instead plays a major role in bulk electrostatics is the spectral asymmetry, which we discuss next.


\subsection{Spectral asymmetry and net charge density}

The imbalance introduced by the anomalous Landau level leads to the so-called spectral asymmetry (see e.g. Ref. \cite{niemi1984spectral}). For eigenstates $\{\alpha\}$ with energies
$\varepsilon_\alpha$ and wave functions $\psi_\alpha(\mathbf{r})$, the
local spectral asymmetry 
is defined as
\begin{align}
    \eta(\mathbf{r})
    &= \sum_\alpha
\mathrm{sgn}\!\left(\varepsilon_\alpha\right)
    |\psi_\alpha(\mathbf{r})|^2.
    \label{eq:eta_def}
\end{align}
This spectral asymmetry enters directly in the expression for the net areal charge density of a Chern semiconductor, $-e \sigma({\bf r})$.
To see that, we begin with the general definition
\begin{equation}
\label{eq:sigma_gen}
\sigma({\bf r})=\sum_\alpha |\psi_\alpha({\bf r})|^2 f_\alpha -\sigma_{\rm ion}({\bf r}),
\end{equation}
where 
\begin{equation}
\label{eq:fermi_dirac}
f_\alpha=\frac{1}{e^{\beta (\varepsilon_\alpha-\mu)}+1}
\end{equation}
is the equilibrium occupation factor for the state $\alpha$  when the chemical potential is $\mu$,  $\beta=1/(k_B T)$ and $T$ is the temperature.
The first term in the right hand side of Eq. (\ref{eq:sigma_gen}) is the local density of electrons and $\sigma_{\rm ion}$ is the local density of positive background charge.

Using the relation
\begin{equation}
\label{eq:total}
\sum_{\alpha}|\psi_\alpha({\bf r})|^2 = \sum_{\alpha\in c}|\psi_\alpha({\bf r})|^2+\sum_{\alpha\in v}|\psi_\alpha({\bf r})|^2,
\end{equation}
we can mathematically rewrite Eq. (\ref{eq:sigma_gen}) in the form
\begin{equation}
\label{eq:sigma_main}
\sigma({\bf r}) =n({\bf r}) -p({\bf r})-\frac{\eta({\bf r)}}{2}+c({\bf r}),
\end{equation}
where
\begin{align}
\label{eq:npetC_main}
n({\bf r}) &= \sum_{\alpha\in {\rm con}}  |\psi_\alpha({\bf r})|^2 f_\alpha\notag\\
p({\bf r}) & = \sum_{\alpha\in {\rm val}}  |\psi_\alpha({\bf r})|^2 (1-f_\alpha)\notag\\
\eta({\bf r}) &= \sum_{\alpha\in {\rm con}}|\psi_\alpha({\bf r})|^2-\sum_{\alpha\in {\rm val}}|\psi_\alpha({\bf r})|^2\notag\\
c({\bf r}) & =\frac{1}{2}\sum_{\alpha}|\psi_\alpha({\bf r})|^2 -\sigma_{\rm ion}({\bf r}).
\end{align}
Here, $\alpha\in {\rm con}$ denotes conduction band states with $\varepsilon_\alpha>0$, $\alpha\in {\rm val}$ denotes valence band states with $\varepsilon_\alpha<0$, $n({\bf r})$ is the areal density of electrons in the conduction bands, $p({\bf r})$ is the areal density of holes in the valence bands, and $c({\bf r})$ is defined as the net ionized dopant concentration.
Since the dimension of the tight-binding Hamiltonian matrix is finite ($2 N \times 2 N$, $N$ being the number of unit cells in the crystal), all the quantities in  Eq. (\ref{eq:npetC_main}) are finite.
When we adopt a low-energy approximation of the electronic structure, assume that it is spatially uniform (or it varies slowly in the scale of a magnetic length), and ignore the sample boundaries,
Eq. (\ref{eq:npetC_main}) simplifies to (cf. App. \ref{app:charge_density} for the derivation)
\begin{align}
\label{eq:npetC2_main}
n({\bf r}) &= \frac{1}{2\pi  \ell_B^2} \sum_{j\in {\rm con}}  f_j\notag\\
p({\bf r}) & = \frac{1}{2\pi  \ell_B^2} \sum_{j\in {\rm val }}  (1-f_j)\notag\\
\eta({\bf r}) &=  \frac{1}{2\pi  \ell_B^2}\left(\sum_{j\in {\rm con }} 1- \sum_{j\in {\rm val }} 1\right)\notag\\
c({\bf r}) & =\frac{1}{2\pi  \ell_B^2} \frac{1}{2}\sum_j 1 -\sigma_{\rm ion},
\end{align}
where $1/(2\pi \ell_B^2)$ is the number of states per unit area in a bulk Landau level, and the sum over $j$ includes ordinary and anomalous Landau levels in all valleys.
The entire ${\bf r}$-dependence is now contained in $\mu$ and $\sigma_{\rm ion}$.
Equation (\ref{eq:npetC2_main}) will be used below to obtain the electrostatic properties of the junction sufficiently far (compared to the penetration depth of edge states) from the edges and interfaces of the system.


In Eq. (\ref{eq:npetC2_main}), $\eta$ counts the difference in the number of electronic states between the conduction and valence bands, per unit area. At zero magnetic field, there are as many bulk electronic states in the conduction band as there are in the valence band (this is obvious from Eq. (\ref{eq:hBHZ})).
In this case, $\eta({\bf r})=0$ in Eq. (\ref{eq:npetC2_main}), and Eq. (\ref{eq:sigma_main}) reads $\sigma=n-p+c$. 
This is the standard expression for the charge density in conventional semiconductors, $c$ being the difference between ionized acceptor and donor concentrations \cite{sze2021physics}.
When a magnetic field is turned on, $c$ in Eq. (\ref{eq:npetC2_main}) stays unchanged because the total number of electronic states $\sum_j 1$ and the positive background charges $\sigma_{\rm ion}$ are insensitive to the magnetic field. But, in Chern insulators, $\eta$ is generically no longer zero when $B \neq0$.
The reason for it is that the valence or the conduction band hosts an extra ($j=0$) bulk Landau level in a magnetic field. 
Specifically, a single Dirac fermion at high-symmetry point $i$ makes the following contribution to $\eta$:
\begin{equation}
    \eta^{(i)} = -\frac{\chi^{(i)}\,\mathrm{sgn}(M^{(i)} B)}{2\pi\ell_B^2}=-\frac{2 e}{h} {\cal C}^{(i)} B,
    \label{eq:eta_i}
\end{equation}
where $1/(2\pi l_B^2)$ is the number of states per unit area in the anomalous Landau level. 
Summing over the four high-symmetry points then yields
\begin{equation}
    \eta = -\frac{2 e }{h} {\cal C} B,
    \label{eq:eta_chern_main}
\end{equation}
with $\mathcal{C}$ the Chern number of the valence band at zero external magnetic field (Eq. (\ref{eq:chern})).
Therefore, the magnetic-field-induced spectral asymmetry is fixed by the product ${\cal C} B$.

It is well-known  \cite{streda1982theory, tutschku2020temperature} that, at zero temperature and with the chemical potential inside the bulk gap,  the spectral asymmetry gives the quantized Hall conductivity through the St\v{r}eda relation
\begin{equation}
 \lim_{T\to 0}\left(\frac{\partial (- e \sigma)}{\partial B}\right)_\mu =\frac{e}{2} \left(\frac{\partial \eta}{\partial B}\right)_\mu = -\frac{e^2}{h}{\cal C},  
\end{equation}
where we have put $n=p=0$ in Eq. (\ref{eq:npetC2_main}), as appropriate for zero temperature.
In contrast, less appears to be known about how the spectral asymmetry influences the electrostatics of Chern junctions. 
This is the subject of the present work.

\subsection{Carrier densities}
\label{subsec:carriers}

For later reference, we evaluate the carrier densities $n$ and $p$ in a homogeneous, bulk, Chern semiconductor.
We focus on the parameter regime where the smallest bulk gap at zero field,
\begin{equation}
 \Delta\equiv 2 |M^{(l)}|,   
\end{equation}
is located at a particular valley $l$ of partial Chern number ${\cal C}^{(l)}$.
At first, we consider the regime where $|\Delta/2\pm \mu|\gg k_B T$, which is analytically convenient and indeed most relevant in conventional semiconductor junctions.
Then, the carrier densities are dominated by thermal
activation near the band edges at valley $l$.
The number of thermally activated carriers is relatively negligible in the valleys that are not degenerate with $l$, as we assume that the energy gaps therein are significantly (several times the thermal energy $k_B T$) larger than $\Delta$.
With these approximations,
\begin{align}
\label{eq:nG_pG}
n &\simeq\frac{g^{( l)}}{2\pi  \ell_B^2}\sum_{j\in{\rm con}}e^{-\beta(\varepsilon_j^{(l)}-\mu)},\notag\\
p &\simeq\frac{g^{(l)}}{2\pi  \ell_B^2}\sum_{j\in{\rm val}}e^{-\beta(\mu-\varepsilon_j^{(l)})},
\end{align}
where $g^{(l)}$ is the valley degeneracy: $g^{(l)}=1$ for $l\in\{\Gamma, {\cal M}\}$, and $g^{(l)}=2$ if $l=X$ (the carrier densities at $X$ and $Y$ are the same due to the $C_{4 z}$ rotational symmetry of the model, so we only consider one of the two, say $X$, with $g^{(X)}=2$).
Note that, while $n$ and $p$ are mainly determined by the low-energy Landau levels of a given valley, the spectral asymmetry $\eta$ is a global (topological) property of the full Brillouin zone set by the total Chern number. 

In the weak-field regime ($A /\ell_B \ll \{k_BT,\Delta\}$), the Landau level sums in Eq. (\ref{eq:nG_pG}) can be evaluated at
linear order in $B$ using the Euler-Maclaurin formula \cite{sharapov2015thermodynamic}.
Combining Eqs. (\ref{eq:LL_ordinary_dirac}), (\ref{eq:LL_anom_dirac}) and (\ref{eq:nG_pG}), we arrive at
\begin{align}
\label{eq:np_MB_weak0}
    n &\simeq
    g^{(l)}\frac{e^{-\beta(\Delta/2-\mu)}}{4\pi A^2\beta^2}
    \left(\beta\Delta-\frac{2 A^2\beta^2}{\ell_B^2}{\cal C}^{(l)}{\rm sgn}( B)\right),\notag\\
    p &\simeq
    g^{(l)}\frac{e^{-\beta(\Delta/2+\mu)}}{4\pi A^2\beta^2}
    \left(\beta\Delta+\frac{2 A^2\beta^2}{\ell_B^2}{\cal C}^{(l)}{\rm sgn}( B)\right),
\end{align}
to first order in $B$.
The difference in sign between the second terms in the two lines of Eq. (\ref{eq:np_MB_weak0}) originates from the anomalous Landau level, which is located in the conduction band when $C^{(l)} B<0$ and in the valence band when $C^{(l)} B>0$.
Equation (\ref{eq:np_MB_weak0}) can be rewritten in a more informative way as
\begin{align}
    \label{eq:np_MB_weak}
    n &\simeq
    \frac{g^{(l)}}{\lambda_{\rm th}^2} e^{-\beta(\Delta/2-\mu)}
    \left[1-\beta\hbar\omega_c {\cal C}^{(l)}\right],\notag\\
    p &\simeq
    \frac{g^{(l)}}{\lambda_{\rm th}^2} e^{-\beta(\Delta/2+\mu)}
    \left[1+\beta \hbar\omega_c{\cal C}^{(l)}\right],
\end{align}
where
\begin{equation}
\lambda_{\rm th}=\frac{2\pi\hbar}{\sqrt{2\pi m^* k_B T}}
\end{equation}
is the thermal de Broglie wavelength,
\begin{equation}
\omega_c=\frac{e B}{m^*}
\end{equation}
is the cyclotron frequency (note that, in our definition, $\omega_c$ can be positive or negative depending on the sign of $B$), and
\begin{equation}
    m^*=\frac{\hbar^2\Delta}{2 A^2}
\end{equation}
is the effective mass of valley $l$ obtained from Eq. (\ref{eq:hi}) in the absence of a magnetic field. The effective mass, positive definite, is the same for the electrons and holes, as we assume in Eq. (\ref{eq:hBHZ}) a particle-hole symmetric electronic structure at zero magnetic field.
Then, the law of mass action reads
\begin{equation}
\label{eq:loma}
    n p \simeq \left(\frac{g^{(l)}}{\lambda_{\rm th}^2}\right)^2 e^{-\beta \Delta} \equiv n_i^2
\end{equation}
to first order in $B$, where $n_i$ is the intrinsic carrier concentration at $B=0$.

We are able to obtain more accurate analytical results for $n$ and $p$ than the ones in Eqs. (\ref{eq:np_MB_weak}) when we include $O(B_m q_x^2)$ and $O(B_m q_y^2)$ corrections in Eq. (\ref{eq:hi}) first and thereafter in  Eqs. (\ref{eq:LL_ordinary_dirac}) and (\ref{eq:LL_anom_dirac}). 
These improved expressions allow for a better quantitative agreement between the analytical theory and the results obtained numerically Eq. (\ref{eq:hBHZ}) without approximation.
However, since those corrections do not qualitatively affect the main results, we relegate them to Appendix \ref{app:np_details}.

In the strong-field regime\footnote{The ``strong'' magnetic fields considered throughout this work are still weak enough so that (i) the anomalous Landau level does not switch from the conduction to the valence band (or vice versa), and (ii) the magnetic flux traversing a unit cell of the square lattice is much smaller than the quantum of flux.}, the carrier densities are dominated from the lowest conduction band Landau level and the highest valence band Landau level at valley $l$. In this case, we do not resort to the Maxwell-Boltzmann approximation of Eq. (\ref{eq:nG_pG}) and instead work with the Fermi-Dirac distribution to get
\begin{align}
    \label{eq:np_MB_strong}
     n &\simeq
    \frac{g^{(l)}}{2\pi  \ell_B^2}\left(\frac{\delta_{{\rm sgn}({\cal C}^{(l)} B),-1}}{e^{\beta(\varepsilon_0^{(l)}-\mu)}+1}+\frac{1}{e^{\beta(\varepsilon_{1,+}^{(l)}-\mu)}+1}\right) \notag\\
    p &\simeq
     \frac{g^{(l)}}{2\pi  \ell_B^2}\left(\frac{\delta_{{\rm sgn}({\cal C}^{(l)} B),+1}}{e^{\beta(\mu-\varepsilon_0^{(l)})}+1}+\frac{1}{e^{\beta(\mu-\varepsilon_{1,-}^{(l)})}+1}\right),
\end{align}
where $\delta$ is the Kronecker delta and the Landau levels energies are indicated in Eqs. (\ref{eq:LL_ordinary_dirac}) and (\ref{eq:LL_anom_dirac}) (or in Appendix \ref{app:LL_Bm} for better accuracy).
The reason for not using the Maxwell-Boltzmann approximation in the preceding expression is motivated by the fact that, as shown in the next subsection, the chemical potential of a locally charge-neutral Chern semiconductor in the strong field regime approaches, and may even intersect with, the conduction or valence bands\footnote{This is our operational definition of "strong field".}.
Thus the Maxwell-Boltzmann approximation is not applicable there and the Fermi-Dirac distribution must be retained.



\subsection{Chemical potential at charge neutrality}
\label{subsec:chemical_potential_neutrality}

Thus far, we have not specified the value of the chemical potential $\mu$.
However, in a homogeneous bulk Chern semiconductor, we expect
charge neutrality to be satisfied locally \cite{li2007semiconductors}.
Hence, the chemical potential takes a spatially uniform value such that $\sigma=0$ in Eq. (\ref{eq:sigma_main}).

Admittedly, the assumption of local charge neutrality could break down for a finite-size sample at very low temperature and magnetic fields.
In this scenario, the chemical potential would become different at the edge and in the bulk as the magnetic field is turned on. The charge density would be nonzero in the bulk and at the edges, though the system would remain electrically neutral overall \cite{boyanovsky1986physical, bottcher2021fate}. 
Two conditions are required for this configuration to hold.
First, the magnetic field must be very low (tending to zero in the thermodynamic limit $L_x\to \infty$). Indeed, because the ratio of the number of edge states to the number of bulk states scales as $1/L_x$, only a small bulk charge density induced by the magnetic field at a fixed bulk chemical potential can be compensated by the edges.
Second, when the bulk has a finite electrical conductivity $\sigma$ due to thermally excited electrons in the conduction band and holes in the valence band, the charge imbalance between the bulk and the edge relaxes and the local charge neutrality is restored on a time scale $\sim \epsilon/\sigma$, where $\epsilon$ is the dielectric constant. Very low temperatures are needed so that the relaxation time is longer than experimentally relevant timescales.
Despite the preceding caveats, for the rest of the paper we will keep assuming that a homogeneous Chern insulator is locally charge neutral. 

Next, we present analytical expressions for $\mu$ at charge neutrality in a pristine Chern semiconductor, for which $c=0$ in Eq. (\ref{eq:npetC2_main}).
These expressions are obtained by solving the equation
\begin{equation}
\label{eq:neutrality_pristine}
    n-p=-\frac{e}{h}\mathcal{C} B.
\end{equation}
We focus once again in the limits of ``weak'' and ``strong'' magnetic fields, in the situation where $n$ and $p$ are located predominantly in the valley $l$ with the lowest energy gap.


Let us begin with the weak magnetic field regime. 
Combining Eqs. (\ref{eq:neutrality_pristine}) and  (\ref{eq:np_MB_weak}), we arrive at
\begin{equation}
   \mu
    \simeq
    \hbar\omega_c \left({\cal C}^{(l)}-\frac{{\cal C}}{2 g^{(l)}} e^{\beta \Delta/2}\right),
    \label{eq:mu_topo_weak}
\end{equation}
which is valid when $\beta|\mu|\ll 1$. 
The fact that $\mu=0$ when $B=0$ is a reflection of the particle-hole symmetry of the energy spectrum at $B=0$ (recall the comment after Eq. (\ref{eq:mass_k})). 

The first term in Eq. (\ref{eq:mu_topo_weak}), proportional to the partial Chern number ${\cal C}^{(l)}$, is the 
contribution from the anomalous Landau level at the $l$ valley, whereas the second term, proportional to the total Chern number, comes from the global spectral asymmetry.
The first term is exponentially small compared to the second term, whenever $\mathcal{C}\neq 0$, since we assume $\beta\Delta\gg 1$.

The appearance of the factor $e^{\beta\Delta/2}$ in Eq. (\ref{eq:mu_topo_weak}) can be understood as follows. At low
temperature, the carrier densities are exponentially small when the
chemical potential lies near midgap. The effect of a small field-induced spectral
asymmetry must therefore be compensated
by a comparatively large shift of $\mu$ toward one of the band edges.

\begin{figure}[h!]
    \centering
    \includegraphics[width=0.8\linewidth]{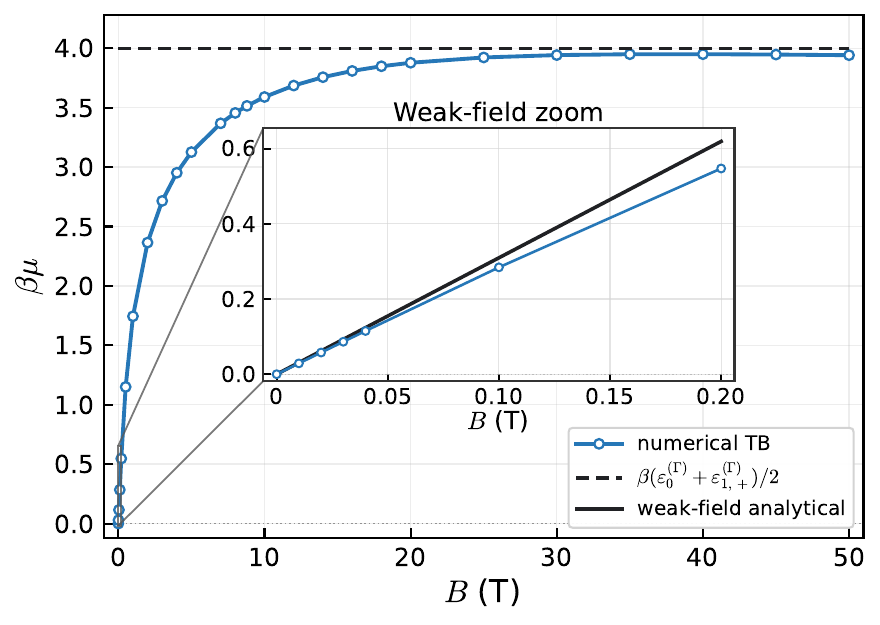}
\caption{
Chemical potential at charge neutrality as a function of magnetic field for a Chern semiconductor with $\mathcal{C}=-1$ and low-energy valley $l=\Gamma$. The parameters are $T\simeq72\,\mathrm{K}$, $A=150\,\mathrm{meV.nm}$, $M=-25\,\mathrm{meV}$ ($\Delta=2|M|=50\,\mathrm{meV}$ and $\beta\Delta=8$), $B_m=-225\,\mathrm{meV.nm^2}$ and $L_x=300\,\mathrm{nm}$. The blue circles are obtained numerically from the full tight-binding Hamiltonian by imposing $\sigma=0$ in Eq.~(\ref{eq:sigma_main}); the calculation includes the full Brillouin zone and therefore accounts for the contributions of all valleys. The inset shows the weak-field regime, where the solid black line corresponds to the analytical result of Eq.~(\ref{eq:mu_topo_weak}) enhanced by the quadratic-in-wave-vector corrections of Appendix \ref{app:np_details}, valid for $\beta|\mu|\ll1$. At stronger magnetic fields, the chemical potential enters the conduction band and approaches the analytical result $\mu\simeq(\varepsilon_0^{(\Gamma)}+\varepsilon_{1,+}^{(\Gamma)})/2$, shown by the dashed line, where $\varepsilon_0^{(\Gamma)}$ and $\varepsilon_{1,+}^{(\Gamma)}$ are the two lowest conduction-band Landau levels at the $\Gamma$ valley.  
For the parameters considered here, $(\varepsilon_0^{(\Gamma)}+\varepsilon_{1,+}^{(\Gamma)})/2=25\,\mathrm{meV}$ is independent of $B$. 
}
    \label{fig:betamu_regimes}
\end{figure}

Let us now move on to the strong field regime, where the chemical potential at charge neutrality approaches, and can eventually intersect with, one of the bulk bands. 
If $\mathcal{C}B<0$, $\mu$ is pushed toward the conduction band and the carrier density is dominated by the two lowest conduction-band Landau levels of the  valley $l$, 
with a negligible hole density in the valence band.
Adopting this two-Landau-level approximation\footnote{Two conduction (valence) band Landau levels need to be included when ${\cal C}B< 0$ (${\cal C} B>0$) in the strong field regime, because $\mu$ tends asymptotically to a value that is halfway between the two (see Eqs. (\ref{eq:mu_strong1}) and (\ref{eq:mu_strong2})). Had we included only one Landau level in the conduction (valence) band, the charge neutrality condition could not be satisfied. In contrast, higher ($j\geq 2$) Landau levels can be neglected when $A/l_B\gg (k_B T,\Delta)$.
} and recalling that the only nonzero value of the Chern number in Eq. (\ref{eq:hBHZ}) has a unit modulus, Eqs. (\ref{eq:neutrality_pristine}) and (\ref{eq:np_MB_strong}) yield
\begin{equation}
\label{eq:mu_strong1}
\mu\simeq\left\{\begin{array}{ll} \left(\varepsilon_0^{(l)}+\varepsilon_{1,+}^{(l)}\right)/2
&\text{if } l=\Gamma, {\cal M}\\
\varepsilon_0^{(l)} & \text{if } l=X, Y
\end{array}\right.,
\end{equation}
with $\varepsilon_0^{(l)}>0$ (anomalous level in conduction band) 
Similarly, if ${\cal C} B>0$, $\mu$ is pushed toward the valence band and the carrier density in the strong field regime is mainly determined by the two highest valence band Landau levels, with a negligible electron concentration in the conduction band. Then, we have
\begin{equation}
\label{eq:mu_strong2}
\mu\simeq\left\{\begin{array}{ll} \left(\varepsilon_0^{(l)}+\varepsilon_{1,-}^{(l)}\right)/2
&\text{if } l=\Gamma, {\cal M}\\
\varepsilon_0^{(l)} & \text{if } l=X, Y
\end{array}\right.,
\end{equation}
with $\varepsilon_0^{(l)}<0$ (anomalous level in valence band).
Finally, if ${\cal C}=0$, the neutrality condition $n=p$ sets 
\begin{equation}
\label{eq:mu_strong3}
    \mu\simeq\text{middle of the bandgap},
\end{equation}
which is no longer at zero energy due the anomalous Landau level that breaks particle-hole symmetry.

Figure~\ref{fig:betamu_regimes} compares the preceding analytical results for $\mu$  with the numerical chemical potential obtained from the full tight-binding model by imposing $\sigma=0$ in Eq.~(\ref{eq:sigma_main}) or in Eq. (\ref{eq:sigma_gen}), with $c({\bf r})=0$. The figure illustrates the case  $\mathcal{C}=-1$, with $l=\Gamma$, $B>0$ and $\beta\Delta=8$. The weak-field analytical approximation matches well with the tight-binding result at low magnetic fields, while at higher fields the numerically calculated chemical potential asymptotically approaches the analytical prediction from Eq. (\ref{eq:mu_strong1}).
If temperature is lowered (increased), the crossover from the weak to the strong field regime, determined by the condition $|\mu|\sim k_B T$, takes place at lower (higher) magnetic fields. In particular, at zero temperature, the chemical potential of a bulk ${\cal C}\neq 0$ Chern insulator moves abruptly to the bulk conduction or valence band as soon as a magnetic field is turned on, rendering the system effectively metallic. The latter observation is in agreement with related remarks in the literature \cite{xiao2010berry, araujo2014chern}.

\section{Poisson equation for a Chern junction}
\label{sec:poisson}


Having discussed the main relevant aspects of a single Chern semiconductor, we now turn to Chern junctions. The electrostatic potential of the Chern junction in Fig. \ref{fig:junction} is obtained by solving the  Poisson equation 
\begin{equation}
    \left(\partial_x^2+\partial_z^2\right)\phi(x,z)
    = e\frac{\sigma(x)}{\epsilon}\,\delta(z)
    \label{eq:poisson_2d_sheet}
\end{equation}
on a finite rectangular domain of dimensions $L_x$ and $L_z$ in the $x$ and in the $z$ directions, respectively.
In the numerical implementation, only the in-plane coordinate $x$ is discretized (step size $\Delta x$) on a uniform grid, while the transverse dependence in $z$ is treated analytically as explained below. For the calculations, we use $L_x=20~\mu\mathrm{m}$, $\Delta x=2~\mathrm{nm}$ (thus $N_x=10001$ grid points) and $L_z=40~\mu\mathrm{m}$.

In Eq. (\ref{eq:poisson_2d_sheet}), dielectric constant $\epsilon$ is assumed to be homogeneous throughout the system\footnote{This assumption, based on simplicity, is justifiable in junctions formed out of one single type of material (homojunctions) that can realize different values of the Chern number through tuning of the low-energy electronic bands. Then, the dielectric constant $\epsilon$ originating from electronic bands remote from the Fermi level can be assumed to be uniform. We will likewise ignore any magnetic field-dependence of $\epsilon$, as remote bands will be affected little by the modest magnetic fields considered in the present work. In case of junctions made from different materials (heterojunctions), $\epsilon$ should take different values on the right and on the left side of the junction. At any rate, the main results discussed below can be readily adapted to this situation.}and $-e\sigma\delta(z)$ is the net 3D charge density, where $\delta(z)$ is the Dirac delta\footnote{The Dirac delta $\delta(z)$ is treated analytically through the discontinuity of the normal electric field across the charged sheet and therefore does not require any transverse discretization or numerical regularization.}. The electrostatic potential $\phi(x,z)$ depends on the in-plane coordinate $x$ and the out-of-plane coordinate $z$, but not on $y$ (due to translational symmetry). 
Even though the charge is confined within the 2D material and the junction is formed along the $x$ axis, a one-dimensional form of the Poisson equation, such as the one used in textbook treatments of conventional 3D junctions \cite{sze2021physics},  is not appropriate here because  electric field lines extend also out of the plane.

To numerically solve Eq. (\ref{eq:poisson_2d_sheet}),  $\sigma$ must be expressed in terms of the electrostatic potential $\phi$.
This is done by recognizing that, in thermodynamic equilibrium, the electrochemical potential $\mu-e\phi$ is spatially uniform. As a result, the chemical and electrostatic potentials are related by $\mu(x) = e\phi(x,0)$, modulo an arbitrary constant that we set to zero. Replacing this relation in Eqs. (\ref{eq:sigma_main}), (\ref{eq:npetC_main}) and (\ref{eq:fermi_dirac}), we have the desired expression for $\sigma(x)$ in terms of $\phi(x,0)$.

\subsection{Boundary conditions}
\label{subsec:bc}

Boundary conditions must be chosen in the $x$ and $z$ directions in order to solve the Poisson equation.
We introduce ohmic contact regions near the left and right boundaries of the system ($x=\pm L_x/2$). These contacts pin the electrostatic potential along the entire ($z\in[-L_z/2, L_z/2])$ lateral boundaries according to
\begin{equation}
    \phi(x,z)=\left\{\begin{array}{ll} \phi_{\mathrm L},  &x\in[-L_x/2,-L_x/2+l_x] \\
    \phi_{\mathrm R},  &x\in[L_x/2-l_x,L_x/2] \end{array}\right.,
    \label{eq:lateral_bc}
\end{equation}
where $\phi_{\mathrm L}$ and $\phi_{\mathrm R}$ are the electric potentials obtained by imposing $\sigma=0$ at the left and right boundaries, and the length $l_x$ describes the spatial overlap between the contact pad and the 2D material.
In a magnetic field, $\phi_L$ and $\phi_R$ are functions of the Chern numbers ${\cal C}_L$ and ${\cal C}_R$, respectively, because of the spectral asymmetry contribution to $\sigma$ in Eq. (\ref{eq:sigma_main}).
We assume $l_x$ to far exceed the decay length of the topological edge states at $x=\pm L_x/2$ into the bulk. Accordingly, those edge states will play no role in the 2D electrostatic properties of the junction. 

The boundaries of the system at $z=\pm L_z/2$ are artificial, introduced to truncate the out-of-plane electrostatic domain. Therein we impose 
\cite{nipane2018boundary}
\begin{equation}
    \left.
    \frac{\partial \phi}{\partial z}
    \right|_{z=\pm L_z/2}
    =0,
    \label{eq:transverse_bc}
\end{equation}
corresponding to a negligible residual normal electric field sufficiently far from the monolayer. We have verified that, for a sufficiently large transverse size $L_z$, the self-consistent charge and potential profiles in the 2D material are insensitive to the precise boundary condition imposed on these
artificial boundaries.

\subsection{Self-consistent numerical solution}

The path to a numerical solution of Eq. (\ref{eq:poisson_2d_sheet}) starts with an initial guess for $\phi(x,0)$ and proceeds by iterations.
At iteration \(m\), the chemical potential
$\mu_{m}(x)=e\phi_m(x,0)$ is 
used to compute the areal charge density 
$-e\sigma_{m}(x)$ via
\begin{equation}
    \sigma_m(x) = n_m(x)-p_m(x)-\frac{\eta(x)}{2}+c(x),
    \label{eq:rho_numeric}
\end{equation}
where $n_m$ and $p_m$ are functions of $\mu_m$ (Eqs. (\ref{eq:npetC2_main})), while 
\begin{equation}
    \eta(x)=\left\{\begin{array}{ll}
    -(2 e/h) {\cal C}_L B, & \text{if } x<0,\\
    -(2 e/h) {\cal C}_R B, & \text{if } x>0.
    \end{array}\right.
\end{equation}
Poisson's equation (\ref{eq:poisson_2d_sheet}) is then solved with $\sigma_m(x)$ as a fixed source and with the boundary conditions (\ref{eq:lateral_bc}) and (\ref{eq:transverse_bc}). Rather than discretizing the full $x$--$z$ domain, we expand the electrostatic potential in sine modes along $x$ and eliminate the transverse direction analytically, following standard Green-function and spectral approaches for Poisson problems in rectangular geometries \cite{MorandRobert1982,Swarztrauber1974}.

The resulting electrostatic potential is denoted by $\widetilde{\phi}_{m}$. Since the charge density itself depends on the potential, the calculation defines a nonlinear self-consistent fixed-point problem analogous to a Gummel-type iterative scheme \cite{Gummel1964SelfConsistent}. 
To improve numerical stability, the input for the next, $(m+1)$-th, iteration is chosen by 
\begin{equation}
    \phi_{m+1}(x,z) = (1-\alpha)\phi_{m}(x,z) + \alpha\widetilde{\phi}_{m}(x,z),
    \label{eq:mixing_numeric}
\end{equation}
where $0<\alpha<1$ is a mixing parameter. The same iterative procedure is repeated until convergence is reached, that is when both $\phi(x,0)$ and $\sigma(x)$ become stationary from one iteration to the next for all $x$ within a prescribed numerical tolerance. Further details of the numerical implementation are given in Appendix~\ref{app:poisson_numerics}.

\subsection{An analytical solution}
\label{subsec:poisson_an}
In Ref. \cite{yu2016carrier}, Yu {\em et al.} solved Eq. (\ref{eq:poisson_2d_sheet}) analytically for a model of two semi-infinite  metallic sheets joined at $x=0$.
Imposing the the boundary conditions
\begin{equation}
\phi(x\to\pm\infty,z=0)=\pm\frac{V_{\rm bi}}{2},
    \label{eq:yakobson_bc}
\end{equation}
they obtained 
\begin{equation}
    \phi(x,z)=\frac{V_{\rm bi}}{\pi}\arctan\!\left(\frac{x}{z}\right)
    \label{eq:yakobson_phi}
\end{equation}
and
\begin{equation}
    -e\sigma(x)=-2\epsilon\left.\partial_z\phi(x,z)\right|_{z=0^+}
    = \frac{2\epsilon}{\pi}\frac{V_{\rm bi}}{x}.
    \label{eq:yakobson_tail}
\end{equation}
Thus, it was predicted that the charge density profile displays a long algebraic tail,
$\sigma(x)\sim 1/x$, instead of being confined to a narrow depletion
region as in a conventional 3D junction. 

In an actual semiconductor or semimetal homojunction with a net ionized dopant concentration $c$ at $x>0$ and $-c$ at $x<0$, the charge density cannot grow
without bound near the interface and the electric potential must be continuous therein. 
More precisely, 
$|\sigma|$ cannot exceed $|c|$ \cite{yu2016carrier}. This constraint truncates the singularity of Eq.~(\ref{eq:yakobson_tail}) at $|x|\lesssim W$, where
\begin{equation}
    W \equiv -\frac{2 \epsilon}{\pi e}\frac{V_{\rm bi}}{c}
    \label{eq:W_2d_scaling}
\end{equation}
is a depletion length within which the areal charge density forms a plateau of value $-e c$.
Note that $W>0$, since $V_{\rm bi}$ and $c$ have the opposite sign.
For $|x|\gg W$, the  behavior of $\sigma$ remains governed by Eq.~(\ref{eq:yakobson_tail}). 
These attributes were confirmed by Yu {\em et al.} from the full numerical solution of the Poisson equation.

The theory of Yu {\em et al.} was conceived for systems with ${\cal C}=0$ and $B=0$. The solution of the Poisson equation for a Chern junction in a magnetic field can be obtained  similarly, albeit with 
two novelties.
First, the product ${\cal C} B$ comes to influence the built-in potential $V_{\rm bi}$.
We elucidate this point in Sec. \ref{sec:built_in}.
Second,
$\sigma$ is given by Eq. (\ref{eq:sigma_main}) instead of by the usual relation $\sigma=n-p+c$. This difference amounts to replacing the $c$ parameter of Yu {\em et al.}'s theory by an effective doping parameter:
\begin{equation}
\label{eq:Ceff}
    c\to c_{\rm eff}=c-\frac{\eta}{2}=c+\frac{e}{h} {\cal C} B,
\end{equation}
which can be nonzero even in a pristine system
\footnote{The topological contribution to $c_{\rm eff}$ has a numerical value $\sim 2.4 \times 10^{10}\, {\rm cm}^{-2}\times {\cal C} B[{\rm T}]$, which need not be small compared to the intrinsic carrier density $n_i$ defined in Eq. (\ref{eq:loma}). For example, $n_i\simeq 2\times 10^{9} {\rm cm}^{-2}$ for the parameter values of Fig. \ref{fig:betamu_regimes}.
For a pristine system ($c=0$), the sign of $c_{\rm eff}$ can be inverted by reversing the direction of the external magnetic field.}. 
In Sec. \ref{sec:charge_density_main}, we will prove through a numerical solution of Eq. (\ref{eq:poisson_2d_sheet}) that the results of Yu {\em et al.} can be transferred to Chern junctions, with the proviso that $c\to c_{\rm eff}$ and that ${\cal C} B$ be incorporated in $V_{\rm bi}$. 



\section{Built-in potential of a pristine Chern junction}
\label{sec:built_in}

Sufficiently far from the interface between two pristine Chern insulators, local charge neutrality is restored, i.e. $\sigma=0$.
This condition determines the chemical potentials $\mu_L$ and $\mu_R$ at the left and right extremities of the system\footnote{In our finite-size numerical simulation, we impose $\sigma=0$ through a boundary condition that mimics ohmic contacts.}.
Then, the built-in potential is defined as
\begin{equation}
    V_{\mathrm{bi}}
    = \frac{\mu_R-\mu_L}{e}=\phi_R-\phi_L.
    \label{eq:Vbi_def}
\end{equation}

Let us begin by considering the conceptually simplest junction, the homojunction, in which (i) the low-energy fermions are located at the same valley $l$ on both sides of the junction (thus $g^{(l)}\equiv g$ is the same), and (ii) $A$ and $\Delta$  are identical for the two sides, but ${\rm sgn}(M^{(l)})$ can be opposite (if valley $l$ undergoes a band inversion at the junction interface). Then, in the weak magnetic field regime, Eq.~\eqref{eq:mu_topo_weak} yields
\begin{equation}
    e V_{\mathrm{bi}}
   \simeq
    \frac{\hbar\omega_c}{2 g}
    \left(2 g-e^{\beta\Delta/2}\right)({\cal C}_R-{\cal C}_L)
    \simeq -\frac{\hbar\omega_c}{2 g}
   e^{\beta\Delta/2}({\cal C}_R-{\cal C}_L) ,
    \label{eq:Vbi_MB}
\end{equation}
where we have used  $g^{(l)}{\cal C}_R^{(l)}-g^{(l)}{\cal C}_L^{(l)}={\cal C}_R-{\cal C}_L$. 
Equation (\ref{eq:Vbi_MB}), valid when $\beta\Delta\gg 1$ and $\beta e|V_{\rm bi}|\ll1$, singles out the topological contribution to the built-in potential, proportional to the
magnetic field and to the change in the Chern number across the junction. 
It can also be rewritten as
\begin{equation}
\label{eq:Vbi_MB2}
    e V_{\rm bi}\simeq -\frac{k_B T}{4\pi l_B^2 n_i}  ({\cal C}_R-{\cal C}_L) {\rm sgn}(B),
\end{equation}
where $n_i$ is the intrinsic carrier density of Eq. (\ref{eq:loma}), with $n_i l_B^2\gg 1$ in the weak-field regime.
The form of Eq. (\ref{eq:Vbi_MB2}) is identical to what one finds for weakly doped conventional semiconductors junctions, except for the fact that the spectral asymmetry takes the role of the doping parameter. 


In the ``strong'' magnetic field regime, approximate analytical expressions for $V_{\rm bi}$ can be obtained from Eqs. (\ref{eq:mu_strong1})-(\ref{eq:mu_strong3}) and (\ref{eq:Vbi_def}).
In short, for a symmetric junction, $e|V_{\rm bi}|$ can become larger than $k_B T$ and comparable to the  energy gap between the bulk conduction and valence band Landau levels when ${\cal C}_L\neq {\cal C}_R$, while $e|V_{\rm bi}|$ vanishes if ${\cal C}_L={\cal C}_R$.
In a conventional $pn$ junction, the built-in potential in the high-doping regime ($|c_L|,|c_R|\gg n_i$) is $e|V_{\rm bi}|=k_B T\ln (|c_L c_R|/n_i^2)$ \cite{sze2021physics}. Yet, this expression is not applicable to a pristine Chern junction in the strong magnetic field regime. The reason is twofold. First, the Maxwell-Boltzmann approximation fails in a Chern semiconductor in the strong field regime. Second, the conduction and valence band of a conventional semiconductor are little affected by the doping concentration, while the bands of a Chern semiconductor vary significantly under a magnetic field. 



\begin{figure}[h!]
    \centering
    \includegraphics[width=\linewidth]{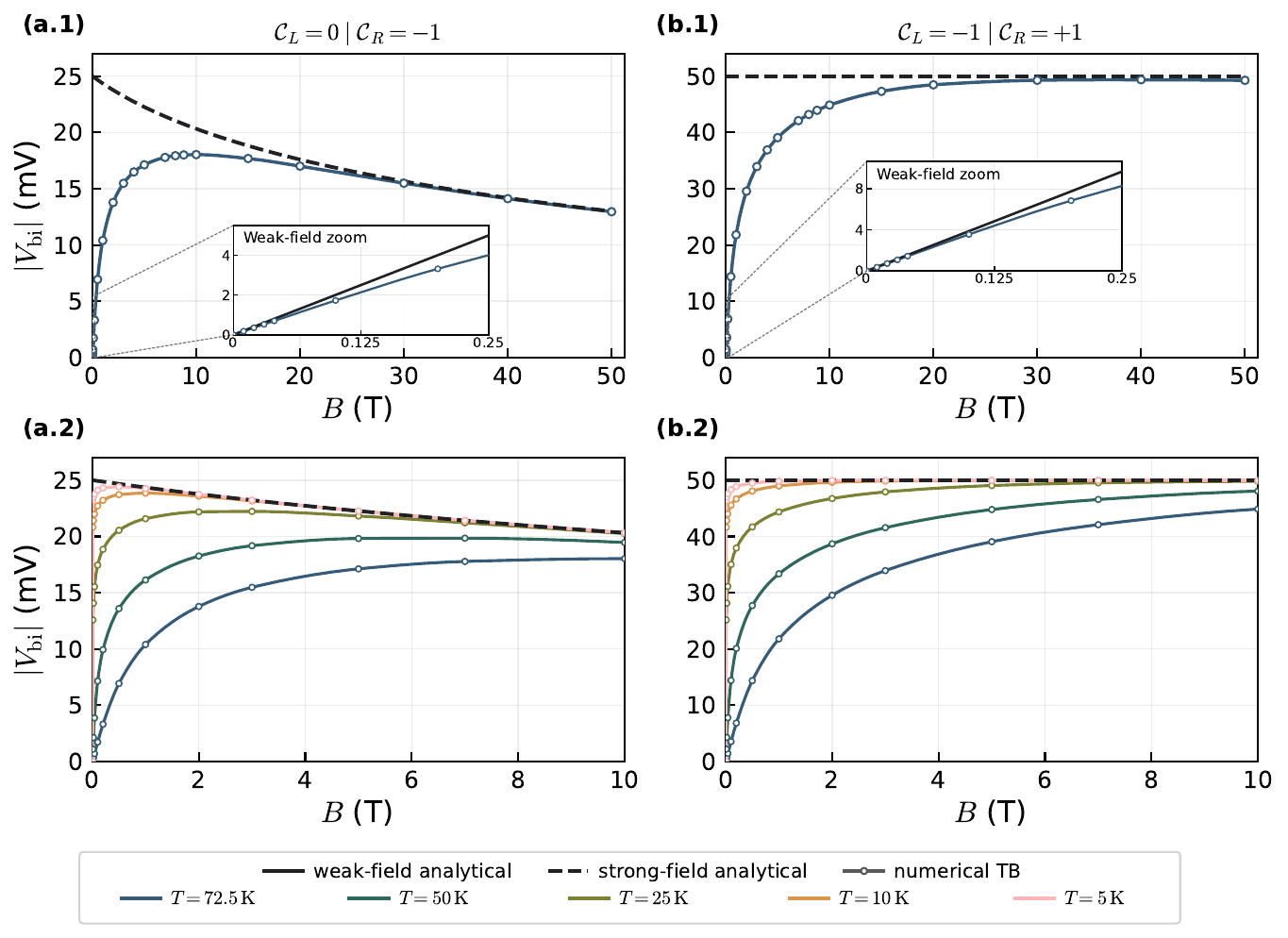}
    \caption{Magnetic-field-induced built-in potential for two Chern junctions: panels (a.1) and (a.2) correspond to $\mathcal{C}_L=0\,|\,\mathcal{C}_R=-1$, while panels (b.1) and (b.2) correspond to $\mathcal{C}_L=-1\,|\,\mathcal{C}_R=+1$. Blue circles in panels (a.1) and (b.1) show the numerical tight-binding results, with $V_{\mathrm{bi}}$ extracted from the difference between the chemical potentials at charge neutrality on the right and left sides of the junction. Panels (a.2) and (b.2) show the temperature dependence of the numerical tight-binding results for $T=72.5$, $50$, $25$, $10$, and $5\,\mathrm{K}$, as indicated by the colors. Black dashed lines denote the strong-field analytical expressions obtained from Eqs. (\ref{eq:mu_strong1})--(\ref{eq:mu_strong3}) and (\ref{eq:Vbi_def}). The insets in panels (a.1) and (b.1) magnify the weak-field regime, $0\leq B\leq0.25\,\mathrm{T}$, and compare the numerical results with the weak-field analytical expression (\ref{eq:Vbi_MB}) enhanced by the quadratic-in-wave-vector corrections of Appendix \ref{app:np_details}. In panels (b.1) and (b.2), the numerical built-in potential saturates close to $|V_{\mathrm{bi}}|=\Delta/e=50\,\mathrm{mV}$. The parameters are $A=150\,\mathrm{meV.nm}$, $B_m=-225\,\mathrm{meV.nm^2}$, $\Delta=50\,\mathrm{meV}$, $a=1\,\mathrm{nm}$, and $L_x=300\,\mathrm{nm}$. For $T=72.5\,\mathrm{K}$, $\beta\Delta=8$. The Dirac-mass parameters are $(M_L,M_R)=(25,-25)\,\mathrm{meV}$ in panels (a.1) and (a.2), and $(-25,-1775)\,\mathrm{meV}$ in panels (b.1) and (b.2).
    }
    \label{fig:built-in}
\end{figure}

A more general junction, a heterojunction, is made up of two semiconductors of different cyclotron frequencies  ($\omega_{c,L}$ and $\omega_{c,R}$), gaps ($\Delta_L$ and $\Delta_R$) and low-energy valley degeneracies ($g_L$ and $g_R$). In this case, Eq. (\ref{eq:Vbi_MB}) in the weak field regime is modified to
\begin{equation}
    e V_{\rm bi}\simeq \frac{\hbar\omega_{c, R}}{2 g_R} \left[2 g_R {\cal C}_R^{(l)}-e^{\beta\Delta_R/2}\mathcal{C}_R\right]-\frac{\hbar\omega_{c, L}}{2 g_L} \left[2 g_L {\cal C}_L^{(l)}-e^{\beta\Delta_L/2}\mathcal{C}_L\right].
    \label{eq:Vbi_MB_general}
\end{equation}
Thus, there is generically a built-in potential induced by a magnetic field
\footnote{In a generic heterojunction one will have a built-in potential even at zero magnetic field, for example if the energy spectrum at zero magnetic field is not particle-hole symmetric. In that case, what we study here is the field-dependent part of the built-in potential.}
even when ${\cal C}_L={\cal C}_R$.
If ${\cal C}_L={\cal C}_R=0$, the built-in potential is still nonzero but exponentially  smaller, for a given (weak) field $B$, compared to the case in which either ${\cal C}_L$ or ${\cal C}_R$ is nonzero. 
If ${\cal C}_L=0\neq {\cal C}_R$, the built-in potential is proportional to ${\cal C}_R$ to leading order.
In a sense, all terms in Eq. (\ref{eq:Vbi_MB_general}) are of band geometric origin (associated to the electronic Berry curvature), even when ${\cal C}_L={\cal C}_R=0$, because of the partial Chern numbers that appear therein.


Figure \ref{fig:built-in} illustrates the order of magnitude of the built-in potential for some reasonable electronic band parameters. Analytical and numerical (tight-binding) results are in good agreement in the weak and strong field regimes. 
As temperature is lowered, the strong field regime is reached at lower  magnetic field strengths\footnote{If $\beta\Delta\gg 1$, the strong field regime (in which the chemical potential at charge neutrality lies near the conduction or valence band edge) can be realized even when $A/l_B\ll k_B T$.}. As a result, at the lowest temperature, the built-in potential can significantly exceed the thermal energy $k_B T$ even at magnetic fields well below $1$ Tesla (see panels (a.2) and (b.2) in Fig. \ref{fig:built-in}) \footnote{We recall that we are assuming that the Chern semiconductor is locally charge neutral sufficiently far from the interface. This assumption could break down in finite size samples at very low temperature and very low magnetic field; see the paragraph above Eq. (\ref{eq:neutrality_pristine}).}.


\section{Charge distribution of a pristine Chern junction}
\label{sec:charge_density_main}

The built-in potential discussed above serves as an input for the self-consistent
numerical solution of the Poisson equation  (Eq. (\ref{eq:poisson_2d_sheet})). The purpose of this section is to determine 
how the electric potential $\phi$  and the net areal charge density $-e\sigma$ vary from one side of the junction to the other. 
This will allow us to compute the junction capacitance in Sec. \ref{sec:capacitance}.

\begin{figure}[h!]
    \centering
    \includegraphics[width=\linewidth]{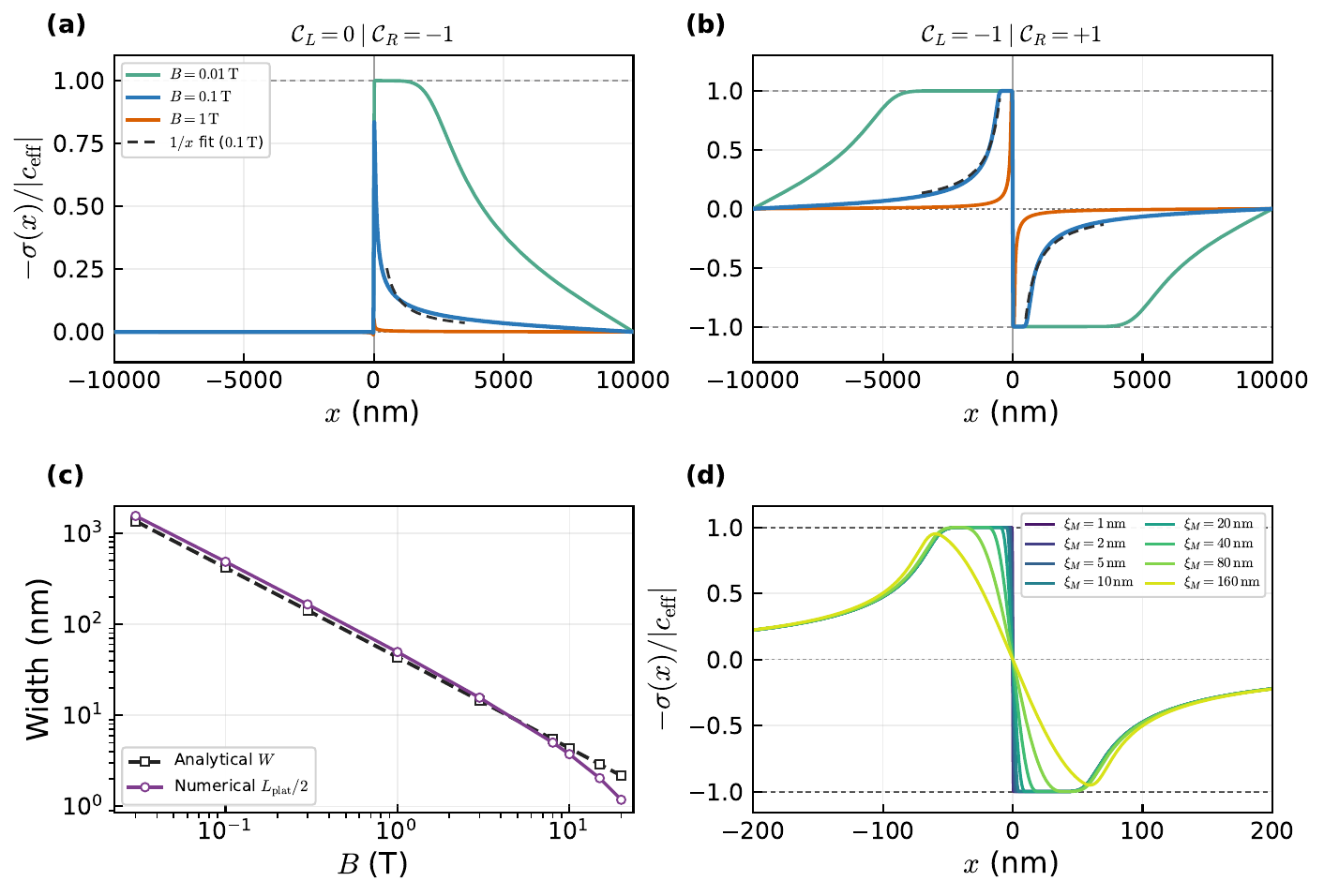}
    \caption{(a) and (b) Charge distribution in two pristine Chern junctions, for different magnetic field strengths and for a temperature of $\sim 7\,\mathrm{K}$. The band inversion near the interface takes place over a length $\xi_M = 40\,\mathrm{nm}$. The band parameters are the same as in Fig.~\ref{fig:built-in} (but notice that the temperature is an order of magnitude lower here). The dielectric constant is $\epsilon=\epsilon_r\epsilon_0$, where $\epsilon_r=6$ and $\epsilon_0$ is the vacuum permittivity. In panel~(a), there are charges (not shown) on the surfaces of the contacts, obtained from $e\sigma_{\mathrm{contact}}(z) = \left. \mp\epsilon\,\partial_x\phi(x,z) \right|_{x=\pm L_x/2}$, where the upper (lower) sign corresponds to the right (left) contact. These charges on the contact surfaces compensate the net charge of the 2D material at $x>0$ and thereby ensure the global charge neutrality of the system. Panel~(c) considers a $\mathcal{C}_L=-1\,|\,\mathcal{C}_R=+1$ junction with $\xi_M=1\,\mathrm{nm}$ and compares the analytical depletion width $W= \frac{2\epsilon}{\pi e} \left| \frac{V_{\mathrm{bi}} }{c_{\mathrm{eff}}} \right|$ (black dashed line) with half of the numerical plateau width, $L_{\mathrm{plat}}/2$ (purple line), as functions of the magnetic field. The numerical plateau is defined as the region satisfying $|\sigma(x)|/|c_{\mathrm{eff}}|\geq 0.99$. Panel~(d) shows the charge profile of the $\mathcal{C}_L=-1\,|\,\mathcal{C}_R=+1$ junction at $B= 1\, \mathrm{T}$ for different mass-variation lengths $\xi_M$. 
    }
    \label{fig:self_consistent_profile}
\end{figure}

The numerical approach to solve the Poisson equation has been summarized in Sec. \ref{sec:poisson} and Appendix \ref{app:poisson_numerics}. A representative result of that calculation is shown for
Fig.~\ref{fig:self_consistent_profile} in pristine homojunctions with  $(\mathcal{C}_L, {\cal C}_R)=(0,1)$ and $(-1,1)$. 
The results of Fig. \ref{fig:self_consistent_profile} are based on a low-energy bulk Landau level description of the electronic structure (with the inclusion of the spatial dependence of the parameter $M$). As such, it neglects chiral edge states an other interfacial states at $x\simeq 0$; those will be included in Sec. \ref{subsec:interface_states} and their impact discussed.

The change in the Chern number across the junction interface is realized through a spatial variation of the mass parameter $M$ over the interval 
$-\xi_M/2<x<\xi_M/2$.
From the point of view of $c_{\rm eff}$, we have an abrupt junction (${\cal C}$ being an integer, $c_{\rm eff}$ changes discontinuously at $x=0$).
But the fact that the Dirac mass changes smoothly over a length $\xi_M$ matters for the charge density profile of the junction. We will return to this point below.

Overall, Fig.~\ref{fig:self_consistent_profile} confirms the traits expected from the analytical treatment of Sec. \ref{subsec:poisson_an}. First, the charge
density is concentrated mainly near the junction interface but it decays slowly (as $1/x$) far from the interface (Fig. \ref{fig:self_consistent_profile} a, b),
in agreement with Eq. (\ref{eq:yakobson_tail}). 
The sign of the charge density is reversed by inverting the sign of the applied magnetic field. 
Second, near the interface, $|\sigma|$  attains a maximum value of $|c_{\rm eff}|$, provided that the temperature is low enough and $\xi_M$ is small enough.
The width $W$ of the charge density plateau is dictated by the ratio between $V_{\rm bi}$ and $c_{\rm eff}$, consistent with Eq. (\ref{eq:W_2d_scaling}); cf. Fig. \ref{fig:self_consistent_profile} c. 
In the weak-field regime ($e |V_{\rm bi}|\ll k_B T$) , $W$ is nearly independent of $B$, as expected from Eq. (\ref{eq:W_2d_scaling}), because both $\eta$ and $V_{\rm bi}$ scale linearly with $B$. Combining Eqs. (\ref{eq:W_2d_scaling}) and (\ref{eq:Vbi_MB}), 
\begin{equation}
\label{eq:W_weak}
    W\sim e^{\beta\Delta/2} a_B/\pi ,
\end{equation}
where $a_B=4\pi\epsilon \hbar^2/(e^2 m^*)$ is the Bohr radius of the Chern semiconductor.
It must be noted that, when temperature is very low compared to the semiconductor gap $\Delta$, the weak field expression (\ref{eq:W_weak}) is attained only for exceedingly small \footnote{
As per Eq. (\ref{eq:mu_topo_weak}), it is required that 
\begin{equation}
 \hbar\omega_c  e^{\beta \Delta/2}\ll k_B T.   
\end{equation}
} values of $B$.

At higher magnetic fields, the built-in potential varies sublinearly with the field (Fig. \ref{fig:built-in}), while $c_{\rm eff}$ still scales linearly with the field.
Thus, $W$ decreases monotonically with $B$ as depicted in Fig. \ref{fig:self_consistent_profile} c.
As $W$ shrinks further and becomes comparable to the lengthscale $\xi_M$ for the variation of the Dirac mass near the interface, the charge plateau begins to disappear (Fig. \ref{fig:self_consistent_profile} d). Indeed, it follows from drawing  semiconductor band diagrams that there cannot be a depleted charge plateau that is narrower than $\simeq \xi_M/2$.  

In sum, the theoretical findings of Ref. \cite{yu2016carrier} are applicable to Chern junctions in a magnetic field, provided that one replaces $c\to c_{\rm eff}$ and provided that the effect of the Chern number is incorporated into the built-in potential. This is one of the main messages of our work.
In Appendix \ref{app:comparison}, we further validate this statement by comparing the charge density distribution for (1) a pristine Chern junction in a magnetic field with an effective doping concentration $c_{\rm eff}$ and (2) a topologically trivial parabolic band semiconductor with the same doping profile in the absence of a magnetic field. The two systems present an identical charge distribution.

\begin{figure}[h!]
    \centering
    \includegraphics[width=0.8\linewidth]{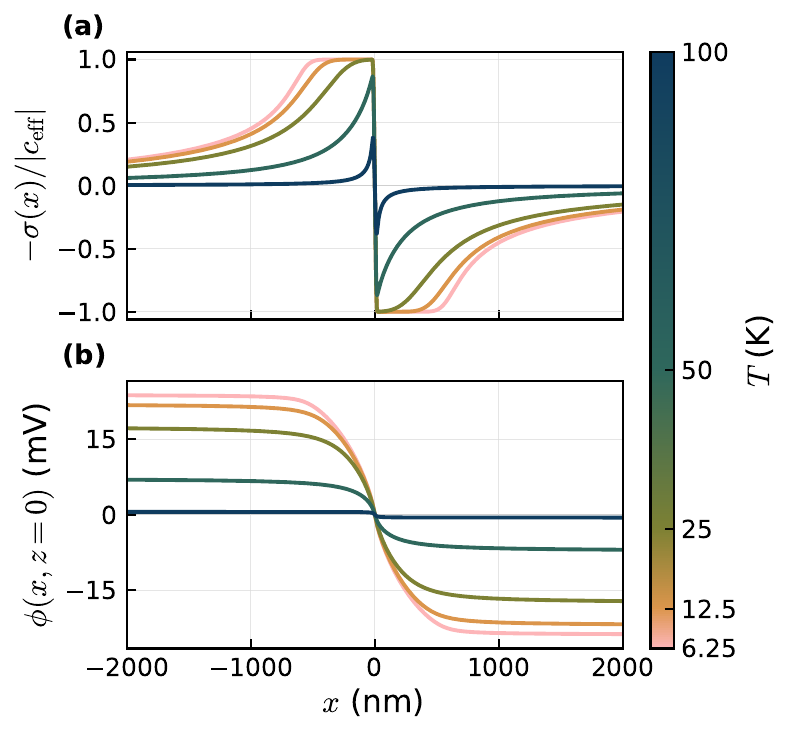}
    \caption{ 
    Influence of temperature dependence on the self-consistent (a) charge density and (b) electrostatic potential profiles of a junction between two Chern-insulating regions with ${\cal C}_\mathrm{L}=-1$ and ${\cal C}_\mathrm{R}=+1$. The magnetic field is $0.1\, \mathrm{T}$ and the band parameters are the same as in Fig. \ref{fig:built-in}.
    In panel(a), the charge depletion plateau shrinks with increased temperature and disappears when $k_B T\gtrsim e|V_{\rm bi}|/2$.
    Panel (b) illustrates the reduction of the built-in potential with increased temperature.
     }
    \label{fig:temperature_dependence}
\end{figure}


In Fig. \ref{fig:self_consistent_profile}, temperature is low enough such that the chemical potential on each side of the junction is close to a semiconductor band edge. 
Figure \ref{fig:temperature_dependence} shows the effect of raising the temperature. First, the built-in potential of topological origin is reduced 
through increased thermal excitation of electrons and holes.  
Specifically, as the intrinsic carrier density $n_i$ exceeds the effective doping $|c_{\rm eff}|$, the chemical potential gradually tends to its intrinsic value  (middle of the semiconductor gap) and consequently the built-in potential of a homojunction is gradually suppressed. 
This trend, which can also be anticipated from Eq. \eqref{eq:Vbi_MB2}, is similar to what takes place in conventional $pn$ homojunctions (see e.g. Fig. C.6 of Ref. \cite{fonstad1994microelectronic}).
Second, much like in conventional 3D junctions \cite{sze2021physics}, the depletion plateau width $W$ of a Chern junction shrinks at higher temperature, as the majority charge carrier concentration in the depletion zone increases.
In Fig. \ref{fig:temperature_dependence}, the depletion plateau is washed out when $2 k_B T\gtrsim e |V_{\rm bi}|$.

One qualitative difference between Chern semiconductors and conventional semiconductors is that $c_{\mathrm{eff}}$ in a pristine Chern junction is independent of temperature and  fixed solely by $\mathcal{C}$ and $B$.
In conventional semiconductors, the concentration of ionized impurities $c$ varies with temperature \cite{sze2021physics}.
In particular, at low temperature, impurities are no longer ionized and the carrier density drops exponentially in the so-called ``freeze-out'' \cite{sze2021physics}.
There is no freeze-out in a pristine Chern semiconductor, where the carrier density fixed by ${\cal C}B$ will remain nonzero even in the zero temperature limit.
As a result, the built-in potential and the charge depletion plateau resulting from a Chern number gradient are present, and indeed prominent, at low temperature (modulo  the caveat of the beginning of Sec. \ref{subsec:chemical_potential_neutrality}).

\subsection{Influence of interface and edge states}
\label{subsec:interface_states}

The bulk Landau-level description of the electronic structure adopted in Fig. \ref{fig:self_consistent_profile} 
does not fully capture the charge redistribution in the  vicinity of the junction interface. Indeed, when the Chern number changes across the interface, chiral interface states emerge inside the bulk gap. These states, not included in the bulk Landau level spectrum, provide additional low-energy electronic degrees of freedom and can therefore modify the local electrostatic response close to $x=0$.

To capture the contribution of the topological edge states without having to combine the tight-binding electronic structure and the Poisson equation over the entire system, we use a numerically less costly hybrid description. The tight-binding and bulk Landau-level descriptions are connected through a smooth window function centered at the interface,
\begin{equation}
w_{\mathrm{TB}}(x)=\frac{1}{2}\left[\tanh\left(\frac{x+W_{\mathrm{TB}}/2}{\lambda}\right)-\tanh\left(\frac{x-W_{\mathrm{TB}}/2}{\lambda}\right)\right],
\label{eq:w_tb}
\end{equation}
where $W_{\mathrm{TB}}=600\,\mathrm{nm}$ is the width within which the full tight-binding electronic structure is used, and $\lambda=5\,\mathrm{nm}$ controls the smoothness of the crossover to the bulk Landau level description\footnote{For numerical efficiency, we truncate the exponentially small tails according to
\begin{equation}
w_{\mathrm{TB}}(x)<10^{-4}\quad\Longrightarrow\quad w_{\mathrm{TB}}(x)=0.
\label{eq:w_tb_cutoff}
\end{equation}
This truncation removes only negligible contributions far from the interface and does not affect the crossover region.}. The numerical value of $W_{\rm TB}$ is chosen to be greater than $\xi_M$, but is otherwise rather arbitrary. The window function approaches unity in the central region and smoothly vanishes away from the interface. Similar hyperbolic-tangent window functions are used to isolate a finite interval while avoiding an abrupt cutoff \cite{djukanovic2024electromagnetic}. 

\begin{figure}[h!]
\centering
\includegraphics[width=0.8\linewidth]{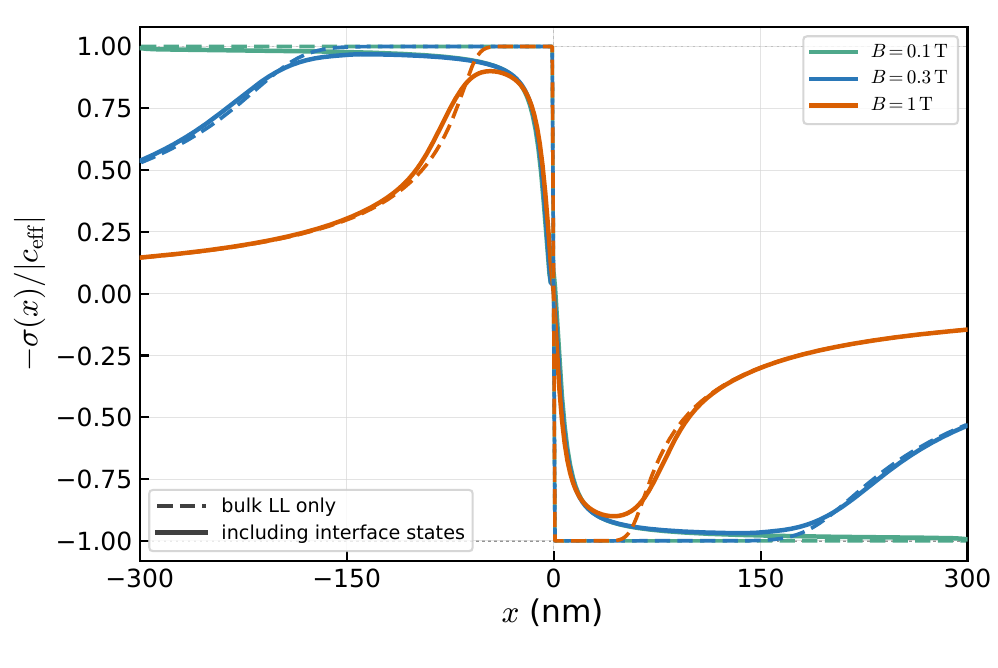}
\caption{
Influence of interfacial electronic states on the self-consistent charge profile of a Chern junction with $\mathcal{C}_L=-1$ and $\mathcal{C}_R=+1$, at $T=6.25\,\mathrm{K}$ and for magnetic fields $B=0.1$, $0.3$, and $1\,\mathrm{T}$. The transition is driven by band inversions at the $X$ and $Y$ valleys, with $(M_L,M_R)=(-875,-925)\,\mathrm{meV}$ and a mass-variation length $\xi_M=1\,\mathrm{nm}$. The other band parameters are $A=150\,\mathrm{meV.nm}$, $B_m=-225\,\mathrm{meV.nm^2}$, and $\Delta=50\,\mathrm{meV}$. Dashed curves are obtained using the bulk Landau-level energy spectrum throughout the junction, whereas solid curves include the tight-binding description of the interface states. Both sets of charge profiles are determined self-consistently using the same Poisson solver. The tight-binding correction is selected by the window function
Eq. \eqref{eq:w_tb},
with $W_{\mathrm{TB}}=600\,\mathrm{nm}$ and $\lambda=5\,\mathrm{nm}$, and is smoothly connected to the bulk Landau-level description outside the interface region.}
\label{fig:interface_states_hybrid}
\end{figure}

The charge density entering the self-consistent Poisson equation is then defined as
\begin{equation}
\sigma_{\mathrm{hyb}}(x)=\bigl[1-w_{\mathrm{TB}}(x)\bigr]\sigma_{\mathrm{LL}}(x)+w_{\mathrm{TB}}(x)\sigma_{\mathrm{TB}}(x),
\label{eq:sigma_hybrid}
\end{equation}
where $\sigma_{\mathrm{LL}}(x)$ denotes the charge density obtained from the bulk Landau-level description of the electronic structure and $\sigma_{\mathrm{TB}}(x)$ is evaluated from the full tight-binding electronic structure that includes interface states.

As shown in Fig.~\ref{fig:interface_states_hybrid}, the correction caused by the tight-binding approach is significant only in the vicinity of $x=0$, over a  distance consistent with the penetration depth of the interface states into the bulk semiconductors.
There, the quasi one-dimensional chiral electron gas smoothens the variation in charge density and reduces the charge depletion.

This result clarifies the respective roles of bulk topology and interface states. The bulk spectral asymmetry fixes the effective doping and therefore controls the built-in potential and the electrostatic profile over long distances. The interface states do not modify this asymptotic behavior, but they affect the short-distance structure of the charge distribution near $x=0$. 

In a finite sample, additional topological states are localized at $x=\pm L_x/2$. In a realistic device geometry, however, these outer edges would be connected to metallic contacts acting as electronic reservoirs. The edge states would then hybridize with the continuum of metallic states and should not be regarded as isolated low-energy channels of the finite sample. 
As a result, the outer edge states should not modify the charge redistribution around $x=0$, nor the built-in potential determined by the bulk neutrality condition. 

\section{Capacitance of a pristine Chern junction}
\label{sec:capacitance}

As a final example of electrostatic observable, we compute the zero-bias junction capacitance divided by the length $L_y$ of the system along the $y$ direction,
\begin{equation}
    C_J=\frac{1}{L_y}\left|\frac{\partial Q_R}{\partial V_{\rm bi}}\right|,
\end{equation}
where
\begin{equation}
    Q_R = -e L_y  \int_0^{L_x/2} dx \sigma(x) +Q_{\rm 
    R,contact}
\end{equation}
is the total net charge in the $x>0$ region (including the charge $Q_{\rm contact}$ at the surface of the right contact, if present). 
We could have equally well used the total charge on the $x<0$ region, since the system is overall charge neutral. 
\begin{figure}[h!]
\centering
\includegraphics[width=\linewidth]{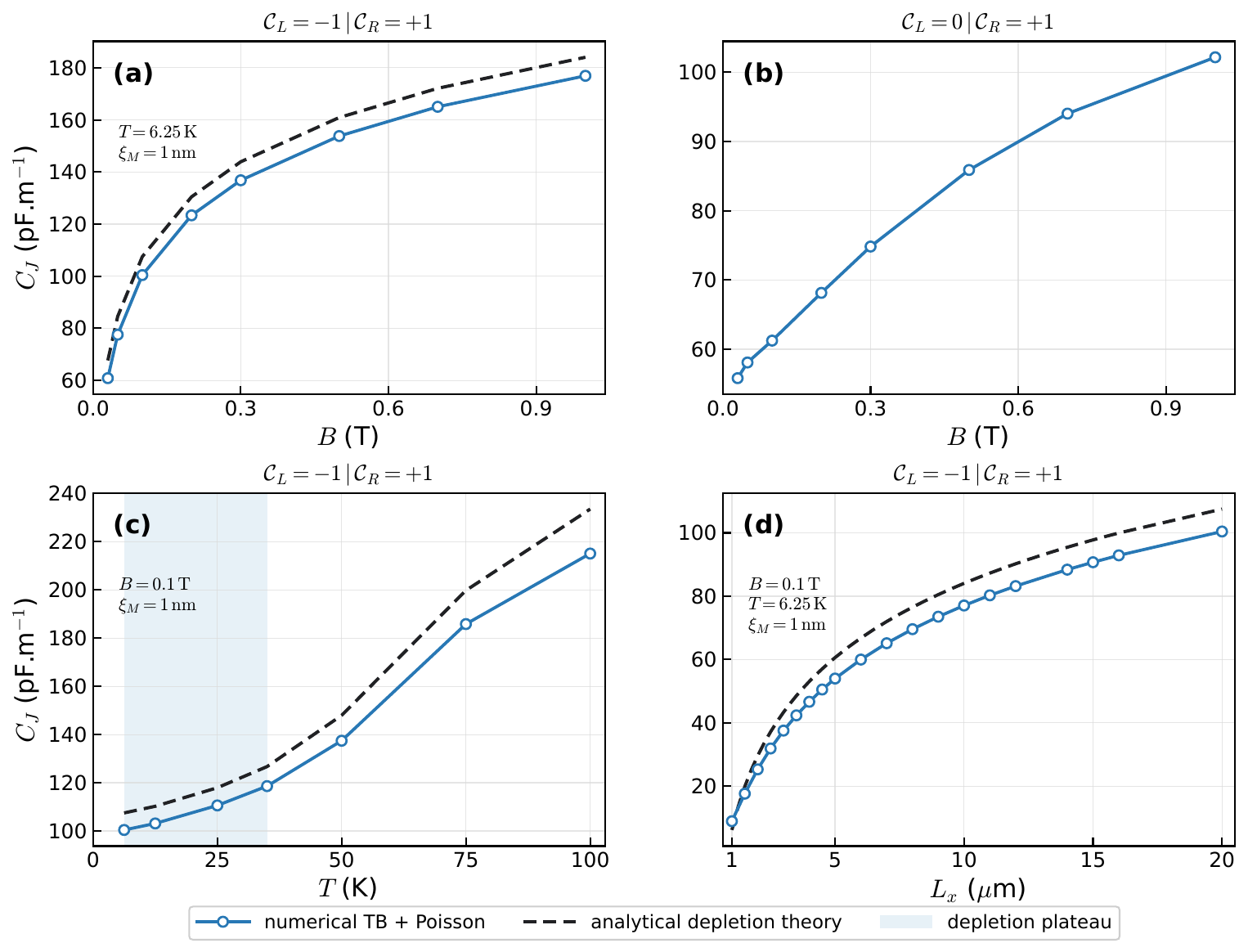}
\caption{Intrinsic capacitance per unit length of Chern junctions, obtained from the self-consistent finite-field tight-binding equation of state combined with the Poisson solver. Blue circles denote the numerical results, while black dashed lines show the analytical depletion expression $C_J^{\mathrm{an}}=(2\epsilon/\pi)\ln[L_x/(2W)]$, with $W=(2\epsilon/\pi e)|V_{\mathrm{bi}}/c_{\mathrm{eff}}|$ and $V_{\rm bi}$ obtained numerically as explained in Sec. \ref{sec:built_in}. (a) Magnetic-field dependence for the symmetric junction $\mathcal{C}_L=-1\,|\,\mathcal{C}_R=+1$. (b) Corresponding numerical result for the asymmetric junction $\mathcal{C}_L=0\,|\,\mathcal{C}_R=+1$; the symmetric depletion formula is not applied in this case. (c) Temperature dependence of the symmetric junction at $B=0.1\,\mathrm{T}$. The shaded region indicates the temperatures for which a depletion plateau is present, defined by $|\sigma(x)|/|c_{\mathrm{eff}}|\geq0.99$. The analytical expression for $C_J$ follows qualitatively the numerical result as a function of temperature, with a worsening agreement at higher temperature. (d) Dependence on the junction length $L_x$ for the symmetric junction at $B=0.1\,\mathrm{T}$ and $T=6.25\,\mathrm{K}$. Unless varied explicitly, the parameters are $A=150\,\mathrm{meV.nm}$, $B_m=-225\,\mathrm{meV.nm^2}$, $\Delta=50\,\mathrm{meV}$, $\xi_M=1\,\mathrm{nm}$, $\epsilon_r=6$, and $L_x=20\,\mu\mathrm{m}$. The Dirac-mass parameters are $(M_L,M_R)=(-25,-1775)\,\mathrm{meV}$ for the symmetric junction and $(M_L,M_R) = (25,-1775)\,\mathrm{meV}$ for the asymmetric junction.}
\label{fig:differential_capacitance}
\end{figure}

For a symmetric $({\cal C}_L=-{\cal C}_R\equiv -{\cal C}$) Chern homojunction without contact charges, the approximate analytical expressions of Sec. \ref{subsec:poisson_an} (see also Table \ref{tab:3D_v_2D}), along with $c\to c_{\rm eff}$, lead to  
\begin{equation}
\label{eq:CJ_an}
C_J=\frac{2\epsilon}{\pi}\ln\left(\frac{L_x}{2 W}\right).
\end{equation}
In a pristine junction,
\begin{equation}
\frac{\partial}{\partial V_{\rm bi}} \exp\left(-\frac{\pi C_J}{2\epsilon}\right) = -\frac{4 \epsilon}{L_x \pi}\frac{h}{e^2} \frac{1}{{\cal C} B}
\end{equation}
is independent of the potential and gives access to the Chern number.


Figure~\ref{fig:differential_capacitance} compares Eq. (\ref{eq:CJ_an}) with the capacitance obtained from the numerical solution of the Poisson equation. The agreement is good when $L_x\gg W\gg \xi_M$. The analytical expression (\ref{eq:CJ_an}) is no longer applicable when the charge plateau is absent or when the junction is asymmetric (${\cal C}_L\neq -{\cal C}_R$). For completeness, the numerical results for the latter cases are displayed in Fig. ~\ref{fig:differential_capacitance}.

\section{Conclusion}
\label{sec:conclusions}

In summary, we have developed a theory of the built-in potential, the charge-density distribution and the  capacitance of Chern semiconductor junctions in the presence of a perpendicular magnetic field $B$. 
Based on a toy model for the electronic structure and a choice of electrostatic boundary conditions, we have provided analytical insights and numerical results that address the question of how a Chern number ${\cal C}$ influences the bulk electrostatic properties of 2D lateral junctions.
We conclude that the electrostatic properties of Chern junctions are overall equivalent to those of conventional (nontopological) 2D junctions, provided that the following changes are realized:
\begin{enumerate}
\item The ionized dopant concentration $c$ of conventional junctions is replaced by
\begin{equation}
    c_{\rm eff} = c+\frac{e}{h}{\cal C}B,
\end{equation}
where $e{\cal C} B/h$ is the band topological contribution originating from spectral asymmetry.
This contribution gives the means of creating $pn$ type junctions through the application of magnetic fields when ${\cal C}$ has opposite signs on both sides of the interface, without having to add dopants.
\item The built-in potential $V_{\rm bi}$ is augmented to include a topological contribution induced by $B$ and sensitive to the Chern numbers of the two sides of the junction. At low enough temperature, this contribution can be as large as the energy gap of the Chern semiconductor, even when $B$ is practically weak. 
For instance, Fig.~\ref{fig:built-in} shows that, at $T=5\,\mathrm{K}$, the $\mathcal{C}_L=-1\,|\,\mathcal{C}_R=+1$ junction gives $e|V_{\mathrm{bi}}|\simeq46.4\,\mathrm{meV}\simeq0.93\Delta$ at $B=0.01\,\mathrm{T}$.
\item The electrostatic contribution from topological edge states is incorporated in the vicinity of the junction interface. 
\end{enumerate}

Some of the results obtained in the present work are analogous to those found for junctions formed between two topologically trivial 2D electron gases (2DEG) with different electron densities and brought into the quantum Hall regime via a perpendicular magnetic field\footnote{Such quantum Hall junctions predate Chern semiconductor junctions; see e.g. Ref. \cite{williams2007quantum} for an experimental realization in graphene.}.
In particular, charge depletion regions and charge density profiles decaying as $1/x$ at long distances 
are known to occur in quantum Hall insulators with electronic density gradients
\cite{kane1988landau, chklovskii1992electrostatics, armagnat2019self, yu2025visualizing}.
From an electrostatic viewpoint, the spatial gradient of the Landau level filling factor in the 2DEG quantum Hall junction plays a role similar to the gradient of the Chern number in a Chern semiconductor junction, whereas the energy difference between the two Landau levels closest to the chemical potential in a 2DEG plays the role of a semiconductor gap.  
Despite this correspondence, the detailed electrostatics of 2DEG quantum Hall junctions is distinct from that of Chern semiconductor junctions.
Next, we indicate two differences.

First, a Chern semiconductor has an energy gap $\Delta$ at zero magnetic field, whereas a 2DEG does not. 
The ``semiconductor energy gap'' in a 2DEG is identical to the magnetic energy scale (e.g. $\hbar\omega_c$, for a non relativistic 2DEG). 
In the quantum Hall regime, this ``semiconductor energy gap'' is much larger than the thermal energy. 
In contrast, in Chern semiconductors, the magnetic energy scale is decoupled from the semiconductor energy gap. There, the Chern number of the crystal at zero field can lead to prominent $B-$induced built-in potential and charge depletion plateau in the regime where the magnetic field energy scale is small compared to $\Delta$. Such regime does not exist in 2DEG quantum Hall junctions. 
Moreover, the alternating compressible-incompressible strips known to emerge at a quantum Hall junction with a density gradient are absent in the Chern junction when the chemical potential is within the semiconductor gap $\Delta$ on both sides of the junction.

Second, in a 2DEG, the  Chern number ${\cal C}$ is zero unless a magnetic field is turned on, and then it is equal to the number of filled Landau levels (times the sign of $B$).
Thus, ${\cal C}\to -{\cal C}$ when $B\to -B$.
In a Chern semiconductor, the value of ${\cal C}$ is nonzero even at $B=0$, and its value remains unchanged when the magnetic field is turned on (up to a certain strength).
As a result, the $B$-induced built-in potential $V_{\rm bi}$  is qualitatively different in both systems. For instance, $V_{\rm bi}$ is an even function of $B$ in 2DEG quantum Hall junction, but $V_{\rm bi}$ is an odd function of $B$ in a Chern homojunction. 
Moreover, the $B$-induced $V_{\rm bi}$ in a (nonrelativistic) 2DEG quantum Hall junction scales as $\hbar \omega_c$ at low temperature, whereas $V_{\rm bi}$ approaches $\Delta$ in a Chern homojunction.

In sum, Chern semiconductor junctions provide a distinct experimental and theoretical platform in which electrostatics is shaped not only by doping and band alignment, but also by the magnetic field $B$ and by the  topological invariants of the $B=0$ electronic bands. 

Going forward, the idealized calculations of the present paper could be adapted to more realistic electronic structures, to disordered systems, to experiment-specific devices, Chern number profiles and boundary conditions, to other materials like quantum spin Hall insulators \cite{calvo2017interplay} and magnetic transition metal dichalcogenide monolayers \cite{xie2024giant}, or to junctions made from higher dimensional topological magnetoelectric materials \cite{chao2024capacitive}. 
In addition, one could examine the role of electrostatics in dynamical (e.g. plasmon \cite{wu1985charge}) and out-of-equilibrium (e.g. longitudinal transport \cite{sze2021physics}) responses of Chern junctions in magnetic fields. 
It could likewise be interesting to conduct further experimental studies to characterize the role of electronic band topology on electrostatics and magnetotransport in Chern junctions. 
In that regard, the setup of Ref. \cite{zhang2024manipulation} appears intriguing. In Fig. 2b of that paper, a junction between ${\cal C}=-2$ and ${\cal C}=2$ Chern semiconductors is realized at a gate voltage of $42$ {\rm V} and a magnetic field of $0.4$ {\rm T}. According to our theory, changing the magnetic field strength away from $0.4$ {\rm T} while keeping the gate voltage fixed would produce a band bending and a built-in potential of topological origin across the junction.

\section*{Acknowledgements}
The authors thank J. Cayssol, S. Guéron, É. Lantagne-Hurtubise, A. Manchon and F. Piéchon for helpful feedback and stimulating questions.
IG acknowledges notes from R. Côté on the Hermite polynomial integrals in App. \ref{app:charge_density}, discussions with C.-C. Chen and M. Holbrook on experimental aspects, and a question from P. Hawrylak informing the author about the compressible-incompressible strips in quantum Hall systems. RD thanks A. Lacerda-Santos for insightful discussions on self-consistent solution schemes in semiconductor devices \cite{lacerda2026a, lacerda2026b}.


\paragraph{Funding information}
RD and IG have been financially supported by the Natural Sciences and
Engineering Research Council of Canada (Grant No. RGPIN-2024-05210). IG benefited
from his affiliation to the Réseau Québécois sur les Matériaux de Pointe, a strategic research network funded by the Fonds de Recherche
du Québec (FRQ) \cite{frqnt}.
All authors have benefited from financial support by the CNRS-Sherbrooke International Research Laboratory on Quantum Frontiers, led by L. Taillefer and C. Proust and funded by the FRQ and the CNRS. 

\begin{appendix}
\numberwithin{equation}{section}

\section{Charge density in a Chern semiconductor}
\label{app:charge_density}
The purpose of this Appendix is to derive Eq.
(\ref{eq:npetC2_main}) from Eq. (\ref{eq:npetC_main}).
To that end, we consider a single massive 2D Dirac fermion, in the $xy$ plane, in the presence of a perpendicular magnetic field pointing along $z$.
We assume that the electronic structure is spatially uniform.  

We first note that the electronic eigenstates can be labelled via $\alpha=(j,X)$, where $j$ is the Landau level index and $X=k_y \ell_B^2$ is the guiding center, $\ell_B$ being the magnetic length and $k_y$ being the wave vector component along the translationally invariant ($y$) direction in the 2D plane.
Solving the Schr\"odinger equation for a harmonic oscillator, we have
\begin{equation}
\label{eq:psi2}
|\psi_{j X}({\bf r})|^2 = u_{j}^2 |h_{j-1, X}({\bf r})|^2+v_{j}^2 |h_{j, X}({\bf r})|^2,
\end{equation}
where ${\bf r}=(x,y)$, $u_j$ and  $v_j$ are real coefficients (crucially independent of $X$) representing the weights of the two components of the Landau level spinors, with $u_{j}^2+v_{j}^2=1$ (see e.g. Refs. \cite{goerbig2009quantum} and \cite{sherafati2019hall}). Also,
\begin{equation}
\label{eq:LLwf}
h_{j, X}(x,y)\equiv \frac{(-1)^j}{\sqrt{L_y}} e^{-i \frac{X y}{\ell_B^2}} \left(\frac{1}{\pi \ell_B^2}\right)^{1/4} \frac{1}{\sqrt{2^j j!}}H_j\left(\frac{x-X}{\ell_B}\right) e^{-\frac{(x-X)^2}{2 \ell_B^2}}
\end{equation}
is a Landau-gauge wave function, where $L_y$ is the length of the system along $y$ and $H_j(x)$ is the $j$-th Hermite polynomial.

Let us now consider the sum
\begin{equation}
\sum_X |h_{j, X}({\bf r})|^2 = \frac{1}{L_y}\left(\frac{1}{\pi \ell_B^2}\right)^{1/2}\frac{1}{2^j j!} \sum_X\left[H_j\left(\frac{x-X}{\ell_B}\right)\right]^2 e^{-\frac{(x-X)^2}{\ell_B^2}}.
\end{equation}
This can be approximated as
\begin{equation}
\sum_X |h_{j, X}({\bf r})|^2  \simeq \int dX \rho(X)  |h_{j, X}({\bf r})|^2 ,
\end{equation}
where $\rho(X)$ is the density of states in $X$. Since $X=k_y \ell_B^2$, we have
\begin{equation}
\rho(X) = \rho(k_y) \frac{d k_y}{dX} = \frac{L_y}{2\pi} \frac{1}{\ell_B^2},
\end{equation}
so that
\begin{align}
\sum_X |h_{j, X}({\bf r})|^2  &\simeq \frac{1}{2\pi \ell_B^2}\left(\frac{1}{\pi \ell_B^2}\right)^{1/2}\frac{1}{2^j j!}\int dX \left[H_j\left(\frac{x-X}{\ell_B}\right)\right]^2 e^{-\frac{(x-X)^2}{\ell_B^2}}\notag\\
&\simeq \frac{1}{2\pi \ell_B^2}
\left(\frac{1}{\pi \ell_B^2}\right)^{1/2}
\frac{1}{2^j j!}\int dX \left[H_j\left(\frac{X}{\ell_B}\right)\right]^2 e^{-\frac{X^2}{\ell_B^2}}\notag\\
&=\frac{1}{2\pi \ell_B}\left(\frac{1}{\pi \ell_B^2}\right)^{1/2}\frac{1}{2^j j!}\int d\tilde{X} \left[H_j\left(\tilde{X}\right)\right]^2 e^{-\tilde{X}^2}.
\end{align}
In the second equality, we have extended the integral over $X$ from $-\infty$ to $\infty$ (this is justified for bulk states, whose wave functions  (\ref{eq:LLwf}) are exponentially localized in $x$ far from the edges).
Then, the integral in the last line is analytically solvable \cite{gradshteyn2014table}:
\begin{equation}
\sum_X |h_{j, X}({\bf r})|^2  =\sum_X |h_{j-1, X}({\bf r}))|^2 =\frac{1}{2\pi \ell_B^2},
\end{equation}
independent of ${\bf r}$ and $j$.
Accordingly, from Eq. (\ref{eq:psi2}), we have
\begin{equation}
\sum_X |\psi_{j, X}({\bf r})|^2  =\frac{1}{2\pi \ell_B^2}.
\end{equation}
Substituting this relation in Eq. (\ref{eq:npetC_main}) and recalling that the energy eigenvalues are degenerate in the guiding center index ($\varepsilon_{j,X}=\varepsilon_j$), we have 
\begin{align}
\label{eq:npetC2}
n({\bf r}) &= \frac{1}{2\pi \ell_B^2} \sum_{j\in {\rm con}}  f_j\notag\\
p({\bf r}) & = \frac{1}{2\pi \ell_B^2} \sum_{j\in {\rm val}}  (1-f_j)\notag\\
\eta({\bf r}) &=  \frac{1}{2\pi \ell_B^2}\left(\sum_{j\in {\rm con}} 1- \sum_{j\in {\rm val}} 1\right)\notag\\
c({\bf r}) & =\frac{1}{2\pi \ell_B^2} \frac{1}{2}\sum_j 1 -\sigma_{\rm ion}.
\end{align}
This is Eq. (\ref{eq:npetC2_main}) of the main text. 
Though we stated that we were assuming a spatially uniform electronic structure, Eq. (\ref{eq:npetC2_main}) is  still valid if the spatial variations of the electronic structure are slow on the scale of $\ell_B$.
\section{Corrections beyond the linear Dirac approximation}
\subsection{Landau levels including the quadratic $B_m$ correction}
\label{app:LL_Bm}

In the main text, the Landau-level spectrum was first discussed within the linear Dirac approximation. Here, we retain the leading quadratic-in-wave-vector correction originating from the $B_m$ term of the half-BHZ Hamiltonian. The influence of such momentum-dependent mass terms on the Landau-level spectrum of BHZ and Chern-insulator models near the $\Gamma$ point has been discussed, for instance, in Refs.~\cite{bottcher2019survival,bottcher2020fate,bottcher2021fate}.
Here, we reproduce the results according to our sign conventions, and extend them to other high-symmetry points as well.

Expanding Eq.~(\ref{eq:hBHZ}) to second order in the momentum $\mathbf{q}$ measured from each high-symmetry point gives
\begin{align}
H_{\Gamma}
&\simeq
A\left(q_x\sigma_x+q_y\sigma_y\right)
+
\left[
M^{(\Gamma)}
-B_m\left(q_x^2+q_y^2\right)
\right]\sigma_z ,
\label{eq:H_Gamma_Bm}
\\
H_X
&\simeq
A\left(-q_x\sigma_x+q_y\sigma_y\right)
+
\left[
M^{(X)}
+B_m\left(q_x^2-q_y^2\right)
\right]\sigma_z ,
\label{eq:H_X_Bm}
\\
H_Y
&\simeq
A\left(q_x\sigma_x-q_y\sigma_y\right)
+
\left[
M^{(Y)}
-B_m\left(q_x^2-q_y^2\right)
\right]\sigma_z ,
\label{eq:H_Y_Bm}
\\
H_{\mathcal M}
&\simeq
-A\left(q_x\sigma_x+q_y\sigma_y\right)
+
\left[
M^{(\mathcal M)}
+B_m\left(q_x^2+q_y^2\right)
\right]\sigma_z .
\label{eq:H_M_Bm}
\end{align}
The Dirac masses $M^{(i)}$ are those introduced in Table~\ref{tab:dirac}. Notice that the quadratic correction is isotropic around $\Gamma$ and $\mathcal M$, whereas it is anisotropic around $X$ and $Y$.

Introducing the kinetic momentum
\begin{equation}
\boldsymbol{\Pi}
=
\mathbf{q}
+
\frac{e}{\hbar}\mathbf{A},
\end{equation}
we define the ladder operators
\begin{equation}
a=
\frac{\ell_B}{\sqrt{2}}
\left[
\Pi_x-i\,\mathrm{sgn}(B)\Pi_y
\right],
\qquad
a^\dagger=
\frac{\ell_B}{\sqrt{2}}
\left[
\Pi_x+i\,\mathrm{sgn}(B)\Pi_y
\right],
\end{equation}
such that
\begin{equation}
\Pi_x^2+\Pi_y^2
=
\frac{2a^\dagger a+1}{\ell_B^2}.
\label{eq:Pi2_ladder}
\end{equation}

At the $\Gamma$ point, for $B>0$, Eq.~(\ref{eq:H_Gamma_Bm}) therefore becomes
\begin{equation}
H_{\Gamma}
=
\begin{pmatrix}
M^{(\Gamma)}
-\dfrac{B_m}{\ell_B^2}\left(2a^\dagger a+1\right)
&
\dfrac{\sqrt{2}A}{\ell_B}a
\\[2mm]
\dfrac{\sqrt{2}A}{\ell_B}a^\dagger
&
-M^{(\Gamma)}
+\dfrac{B_m}{\ell_B^2}\left(2a^\dagger a+1\right)
\end{pmatrix}.
\label{eq:H_Gamma_ladder}
\end{equation}
This Hamiltonian can be diagonalized in the $\{|j-1\rangle, |j\rangle\}$ subspace, where $|j\rangle$ is defined via $a^\dagger a|j\rangle = j |j\rangle$. 
For $j\geq1$, the energy eigenvalues read
\begin{equation}
\varepsilon_{j,\pm}^{(\Gamma)}
=
\frac{B_m}{\ell_B^2}\mathrm{sgn}(B)
\pm
\sqrt{
\left(
M^{(\Gamma)}
-\frac{2jB_m}{\ell_B^2}
\right)^2
+
\frac{2jA^2}{\ell_B^2}
}.
\label{eq:LL_Gamma_Bm}
\end{equation}
The corresponding eigenstates are 
\begin{equation}
\label{eq:eigenstates}
    |\Psi_{j,\pm}\rangle
    =
    \begin{pmatrix}
        u_{j,\pm}|j-1\rangle\\
        v_{j,\pm}|j\rangle
    \end{pmatrix},
\end{equation}
where $u_{j,\pm}$ and $v_{j,\pm}$ are real coefficients that can be readily determined but unnecessary for the present work. 
The anomalous $j=0$ Landau level has an energy
\begin{equation}
\varepsilon_{0}^{(\Gamma)}
=
-\mathrm{sgn}(B)
\left(
M^{(\Gamma)}
-\frac{B_m}{\ell_B^2}
\right),
\label{eq:LL0_Gamma_Bm}
\end{equation}
with an eigenstate
\begin{equation}
\label{eq:eigenstates2}
    |\Psi_{0}\rangle
    =
    \begin{pmatrix}
        0\\
        |0\rangle
    \end{pmatrix}.
\end{equation}

A similar calculation applies around $\mathcal M$. The change of sign of both Dirac velocities with respect to $\Gamma$ does not modify the chirality, since $\chi^{(\mathcal M)}=\chi^{(\Gamma)}=+1$. However, the quadratic contribution to the Dirac mass has the opposite sign. For $j\geq1$, one obtains
\begin{equation}
\varepsilon_{j,\pm}^{(\mathcal M)}
=
-\frac{B_m}{\ell_B^2}\mathrm{sgn}(B)
\pm
\sqrt{
\left(
M^{(\mathcal M)}
+\frac{2jB_m}{\ell_B^2}
\right)^2
+
\frac{2jA^2}{\ell_B^2}
}.
\label{eq:LL_M_Bm}
\end{equation}
The corresponding anomalous level is
\begin{equation}
\varepsilon_{0}^{(\mathcal M)}
=
-\mathrm{sgn}(B)
\left(
M^{(\mathcal M)}
+\frac{B_m}{\ell_B^2}
\right).
\label{eq:LL0_M_Bm}
\end{equation}
When $|B_m|/l_B^2$ is negligible compared to $|M^{(i)}|$, Eqs.~(\ref{eq:LL_Gamma_Bm})--(\ref{eq:LL0_M_Bm}) reduce to Eqs.~(\ref{eq:LL_ordinary_dirac}) and (\ref{eq:LL_anom_dirac}) of the main text. 
Therefore, the quadratic-in-wave-vector term in the half-BHZ Hamiltonian does not alter the structure of the Landau-level problem at $\Gamma$ and $\mathcal M$, but introduces a magnetic-field-dependent correction to both the ordinary and anomalous Landau-level energies.

The situation is qualitatively different around the $X$ and $Y$ points. Indeed,
\begin{equation}
\Pi_x^2-\Pi_y^2
=
\frac{1}{\ell_B^2}
\left(
a^2+a^{\dagger 2}
\right),
\label{eq:Pi_anisotropic}
\end{equation}
such that the quadratic parts of Eqs.~(\ref{eq:H_X_Bm}) and (\ref{eq:H_Y_Bm}) read
\begin{align}
\label{eq:deltaH_XY}
\delta H_X
&=
\frac{B_m}{\ell_B^2}
\left(
a^2+a^{\dagger 2}
\right)\sigma_z ,
\notag\\
\delta H_Y
&=
-\frac{B_m}{\ell_B^2}
\left(
a^2+a^{\dagger 2}
\right)\sigma_z .
\end{align}
Contrary to the number operator $a^\dagger a$ appearing at $\Gamma$ and $\mathcal M$, the ``squeezing'' terms of Eq. (\ref{eq:deltaH_XY}) are not diagonal in the harmonic-oscillator basis; they couple $|j\rangle$ with $|j\pm 2\rangle$.  As a result, the Hamiltonian can no longer be separated into independent  $\{|j-1\rangle, |j\rangle\}$ sectors and thus it cannot be analytically diagonalized. 
However, Eq. (\ref{eq:deltaH_XY}) implies that 
\begin{equation}
    \langle\Psi^{(0)}_{j}|
    \delta H_{X,Y}
    |\Psi^{(0)}_{j}\rangle=0,
\end{equation}
where $|\Psi^{(0)}_{j}\rangle$, of the form given by Eq. (\ref{eq:eigenstates}) or Eq. (\ref{eq:eigenstates2}), are the eigenstates of the Hamiltonian when $\delta H_{X, Y}=0$.
The first-order correction from $\delta H_{X,Y}$ to the Landau-level energies therefore vanishes. 
The leading energy correction consequently appears at second order in $B_m$.

This is in contrast with the $\Gamma$ and $\mathcal M$ valleys, where the quadratic-in-wave-vector term is proportional to $B_m(2a^\dagger a+1)\sigma_z$. This operator has nonzero diagonal matrix elements in the unperturbed Landau-level states and therefore produces corrections already at linear order in $B_m$. Therefore, for sufficiently small values of $B_m$ ($|B_m|/l_B^2\ll |M^{(i)}|$), the quadratic corrections associated with the $X$ and $Y$ valleys are parametrically weaker than those at $\Gamma$ and $\mathcal M$, and their contribution to the carrier densities can be neglected to leading order.

\subsection{Carrier densities}
\label{app:np_details}

The Landau-level expressions derived in the previous subsection can be used to obtain improved analytical expressions for the carrier densities. We focus on the weak-field regime considered in Eq.~(\ref{eq:np_MB_weak}), and assume that the thermally activated carriers are predominantly associated with either the $\Gamma$ or the $\mathcal M$ valley.

We first consider the $\Gamma$ valley. In the Maxwell-Boltzmann regime, the electron density can be written as
\begin{equation}
n^{(\Gamma)}
\simeq
\frac{1}{2\pi\ell_B^2}
\left[
\sum_{j=1}^{\infty}
e^{-\beta\left(\varepsilon_{j,+}^{(\Gamma)}-\mu\right)}
+
\delta_{\mathrm{sgn}(M^{(\Gamma)}B),-1}
e^{-\beta\left(\varepsilon_{0}^{(\Gamma)}-\mu\right)}
\right],
\label{eq:n_Gamma_Bm_sum}
\end{equation}
where the second term accounts for the anomalous Landau level when it belongs to the conduction band.

Introducing the continuous variable $x=2j/\ell_B^2$,
the sum over ordinary Landau levels can be evaluated in the weak-field regime using the Euler-Maclaurin formula,
\begin{equation}
\sum_{j=1}^{\infty}
F\left(\frac{2j}{\ell_B^2}\right)
\simeq
\frac{\ell_B^2}{2}
\int_0^\infty F(x)\,dx
-\frac{1}{2}F(0).
\label{eq:EM_Bm}
\end{equation}

Close to the conduction-band edge, the dispersion entering the integral can be expanded as
\begin{equation}
\sqrt{
\left(M^{(\Gamma)}-B_m x\right)^2+A^2x
}
\simeq
|M^{(\Gamma)}|
+
\frac{
A^2-2M^{(\Gamma)}B_m
}{
2|M^{(\Gamma)}|
}x.
\label{eq:Gamma_edge_Bm}
\end{equation}
Combining the ordinary and anomalous Landau-level contributions, and retaining terms up to linear order in the magnetic field, gives
\begin{align}
n^{(\Gamma)}
&\simeq
\frac{
e^{-\beta\left(|M^{(\Gamma)}|-\mu\right)}
}{
4\pi\beta^2
\left[A^2-2M^{(\Gamma)}B_m\right]
}
\left[
2\beta|M^{(\Gamma)}|
-
\frac{A^2\beta^2}{\ell_B^2}
\mathrm{sgn}\left(M^{(\Gamma)}B\right)
\right],
\label{eq:n_Gamma_Bm}
\\
p^{(\Gamma)}
&\simeq
\frac{
e^{-\beta\left(|M^{(\Gamma)}|+\mu\right)}
}{
4\pi\beta^2
\left[A^2-2M^{(\Gamma)}B_m\right]
}
\left[
2\beta|M^{(\Gamma)}|
+
\frac{A^2\beta^2}{\ell_B^2}
\mathrm{sgn}\left(M^{(\Gamma)}B\right)
\right].
\label{eq:p_Gamma_Bm}
\end{align}
These expressions assume that the conduction- and valence-band extrema remain located at the $\Gamma$ point, which within the quadratic expansion requires $A^2-2M^{(\Gamma)}B_m>0$.
Alternatively, we can write
\begin{align}
n^{(\Gamma)}
&\simeq
\frac{
e^{-\beta\left(|M^{(\Gamma)}|-\mu\right)}
}{
\lambda_{\rm th}^2
}
\left(
1
-
\beta \hbar\omega_c \mathcal{C}^{(\Gamma)}
\right),
\label{eq:n_Gamma_Bm2}
\\
p^{(\Gamma)}
&\simeq
\frac{
e^{-\beta\left(|M^{(\Gamma)}|+\mu\right)}
}{
\lambda_{\rm th}^2
}
\left(
1
+
\beta \hbar\omega_c \mathcal{C}^{(\Gamma)}
\right),
\label{eq:p_Gamma_Bm2}
\end{align}
with appropriately defined cyclotron frequency and thermal wavelengths at the $\Gamma$ valley (N.B.: unlike in Eq. (\ref{eq:np_MB_weak}), here the effective masses entering in the definitions of $\omega_c$ and $\lambda_{\rm th}$ are not identical). 

The calculation at the $\mathcal M$ valley proceeds in the same way. The opposite sign of the quadratic correction to the Dirac mass amounts to replacing
\begin{equation}
A^2-2M^{(\Gamma)}B_m
\longrightarrow
A^2+2M^{(\mathcal M)}B_m.
\end{equation}
One consequently obtains
\begin{align}
n^{(\mathcal M)}
&\simeq
\frac{
e^{-\beta\left(|M^{(\mathcal M)}|-\mu\right)}
}{
4\pi\beta^2
\left[A^2+2M^{(\mathcal M)}B_m\right]
}
\left[
2\beta|M^{(\mathcal M)}|
-
\frac{A^2\beta^2}{\ell_B^2}
\mathrm{sgn}\left(M^{(\mathcal M)}B\right)
\right],
\label{eq:n_M_Bm}
\\
p^{(\mathcal M)}
&\simeq
\frac{
e^{-\beta\left(|M^{(\mathcal M)}|+\mu\right)}
}{
4\pi\beta^2
\left[A^2+2M^{(\mathcal M)}B_m\right]
}
\left[
2\beta|M^{(\mathcal M)}|
+
\frac{A^2\beta^2}{\ell_B^2}
\mathrm{sgn}\left(M^{(\mathcal M)}B\right)
\right].
\label{eq:p_M_Bm}
\end{align}

In the limit $B_m\rightarrow0$, Eqs.~(\ref{eq:n_Gamma_Bm})--(\ref{eq:p_M_Bm}) reduce to Eq.~(\ref{eq:np_MB_weak}) for $\chi^{(\Gamma)}=\chi^{(\mathcal M)}=+1$.

As discussed in the previous subsection, the quadratic $B_m$ term at the $X$ and $Y$ valleys generates squeezing terms that couple different Landau-level sectors. As a consequence, an analogous closed-form derivation of the $B_m$-corrected carrier densities is not available at these valleys. This does not constitute a limitation in the regime $|B_m M^{(i)}|/l_B^2\ll 1$, where the corrections for the carrier densities at $X$ and $Y$ valleys are negligible compared to the corrections at valleys $\Gamma$ and ${\cal M}$.

\section{Numerical solution of the Poisson equation}
\label{app:poisson_numerics}

In this Appendix, we provide additional details on the numerical implementation of the self-consistent Poisson scheme introduced in Sec.~\ref{sec:poisson}. The main simplification is that the transverse coordinate $z$ is treated analytically, so that only the in-plane coordinate $x$ needs to be discretized.

\subsection{Neumann electrostatic kernel}

We decompose the electrostatic potential as
\begin{equation}
    \phi(x,z)=\phi_{\rm lin}(x)+u(x,z),
\end{equation}
where $\phi_{\rm lin}(x)=\phi_L+(x+L_x/2)(\phi_R-\phi_L)/L_x$ satisfies the Dirichlet conditions at $x=\pm L_x/2$. The induced contribution $u(x,z)$ therefore obeys homogeneous Dirichlet conditions at the lateral boundaries.

Introducing $x'=x+L_x/2\in[0,L_x]$, the charge density and induced potential are expanded in sine modes,
\begin{equation}
    \sigma(x)
    =
    \sum_{\ell=1}^{\infty}
    \sigma_\ell\sin(k_\ell x'),
    \qquad
    u(x,z)
    =
    \sum_{\ell=1}^{\infty}
    u_\ell(z)\sin(k_\ell x'),
\end{equation}
with $k_\ell=\frac{\ell\pi}{L_x}$ and $\ell=1,2,\ldots$. In the continuum notation,
\begin{equation}
    \sigma_\ell
    =
    \frac{2}{L_x}
    \int_0^{L_x}
    \sigma(x'-L_x/2)
    \sin(k_\ell x')\,dx'.
    \label{eq:sigma_modal}
\end{equation}
Here, $\ell$ labels the spatial sine modes and should not be confused with the iteration index $m$ used below, e.g. in $\sigma_m(x)$. This separation-of-variables procedure is equivalent to a Green-function treatment of Poisson's equation in a rectangular domain with mixed Dirichlet and Neumann boundary conditions \cite{MorandRobert1982}.

For each mode, the Poisson equation reduces to
\begin{equation}
    -\epsilon \left(\partial_z^2-k_\ell^2\right) u_\ell(z)= -e\sigma_\ell\delta(z).
\end{equation}
For $z\neq 0$, the modal equation is homogeneous. Imposing the Neumann boundary conditions
$\partial_z u_\ell(\pm L_z/2)=0$, together with the continuity of the potential at $z=0$ and the jump condition
\begin{equation}
-\epsilon \left[
\partial_z u_\ell(0^+)-\partial_z u_\ell(0^-)
\right]
=
-e\sigma_\ell,
\end{equation}
gives the full transverse dependence
\begin{equation}
u_\ell(z)
=
-\frac{e\sigma_\ell}{2\epsilon k_\ell}
\frac{
\cosh\left[k_\ell\left(L_z/2-|z|\right)\right]
}{
\sinh\left(k_\ell L_z/2\right)
}.
\label{eq:fullsolution}
\end{equation}
Evaluating Eq.~(\ref{eq:fullsolution}) at the monolayer, $z=0$, gives
\begin{equation}
u_\ell(0)
=
-eG_\ell^{(N)}\sigma_\ell,
\qquad
G_\ell^{(N)}
=
\frac{1}{
2\epsilon k_\ell
\tanh\left(k_\ell L_z/2\right)
}.
\label{eq:kernel}
\end{equation}
In the limit $L_z\rightarrow\infty$, this expression reduces to $G_\ell^{(N)}\rightarrow1/(2\epsilon k_\ell)$, corresponding to a 2D charged sheet in an unbounded transverse space \cite{Krumland2021}.

\subsection{Spectral implementation}

Only the $x$ direction is discretized, with $x_j=-L_x/2+j\Delta x$ and $\Delta x=L_x/(N_x-1)$. For the calculations presented in the main text, we use $L_x=20~\mu{\rm m}$, $\Delta x=2~{\rm nm}$, $N_x=10001$, and $L_z=40~\mu{\rm m}$. No transverse grid or discretization step $\Delta z$ is therefore required.

The charge density on the $N_x-2$ interior points is transformed using a discrete sine transform, which provides the natural discrete sine basis associated with homogeneous Dirichlet boundary conditions at the two lateral boundaries. The discrete transform contains $N_x-2$ sine modes,
$\ell=1,\ldots,N_x-2$,
and all modes resolved by the spatial grid are retained in the calculation. No additional high-$\ell$ cutoff is introduced. The largest resolved wave number is therefore approximately $k_{\rm max}\simeq\pi/\Delta x$.
At large $\ell$, the electrostatic contribution is naturally suppressed by the decay of the modal Green-function kernel, $G_\ell^{(N)}\sim 1/k_\ell$, together with the decrease of the sine coefficients $\sigma_\ell$ for sufficiently smooth charge profiles. Transform-based spectral approaches of this kind provide efficient solvers for separable Poisson and elliptic problems in rectangular geometries \cite{Swarztrauber1974,Gatiso2021,PeiTong2025}.  Each sine component is multiplied by $-eG_\ell^{(N)}$, and an inverse sine transform gives the induced electrostatic potential at $z=0$,
\begin{equation}
    \widetilde{\phi}(x,0) = \phi_{\rm lin}(x) - e\sum_\ell G_\ell^{(N)} \sigma_\ell \sin(k_\ell x').
    \label{eq:appendix_phi_target}
\end{equation}

Although only the value at $z=0$ is required to update the carrier density in the self-consistent loop, the electrostatic potential is known throughout the full transverse domain. Once the modal coefficients $\sigma_\ell$ have been determined, Eqs.~(\ref{eq:fullsolution}) and (\ref{eq:kernel}) give
\begin{equation}
    \widetilde{\phi}(x,z)
    =
    \phi_{\rm lin}(x)
    -
    e\sum_\ell
    G_\ell^{(N)}\sigma_\ell
    \frac{
        \cosh\left[k_\ell\left(L_z/2-|z|\right)\right]
    }{
        \cosh\left(k_\ell L_z/2\right)
    }
    \sin(k_\ell x').
    \label{eq:full_phi_reconstruction}
\end{equation}
Equation~(\ref{eq:appendix_phi_target}) is simply the restriction of this expression to the monolayer, $z=0$. The self-consistency condition is monitored at $z=0$ because the carrier density depends only on $\phi(x,0)$. Once $\phi(x,0)$ and the corresponding coefficients $\sigma_\ell$ have converged, Eq.~(\ref{eq:full_phi_reconstruction}) uniquely determines the converged electrostatic potential for every $z$ in the domain. The full two-dimensional Poisson problem is thus replaced by one-dimensional sine transforms and mode-by-mode multiplication.

\subsection{Self-consistent solution}

For a given electrostatic potential $\phi_m(x,0)$, the local chemical potential $\mu_m(x)=e\phi_m(x,0)$ determines the charge density $\sigma_m(x)$ through Eq.~(\ref{eq:rho_numeric}). Equation~(\ref{eq:appendix_phi_target}) then provides the corresponding updated potential $\widetilde{\phi}_m$.

The self-consistent solution is defined by the nonlinear fixed-point condition
\begin{equation}
F[\phi]
=
\widetilde{\phi}[\phi]-\phi
=
0,
\end{equation}
where $\widetilde{\phi}[\phi]$ denotes the electrostatic potential obtained by first evaluating the charge density from $\phi$ and then solving Poisson's equation. This procedure is analogous to a Gummel-type self-consistent scheme \cite{Gummel1964SelfConsistent}.

To stabilize the fixed-point iteration, the electrostatic potential is updated using linear under-relaxation,
\begin{equation}
\phi_{m+1}(x,0)
=
\phi_m(x,0)
+
\alpha F[\phi_m](x,0),
\end{equation}
or equivalently,
\begin{equation}
\phi_{m+1}(x,0)
=
(1-\alpha)\phi_m(x,0)
+
\alpha \,\widetilde{\phi}_m(x,0),
\label{eq:linearmixing}
\end{equation}
where $0<\alpha<1$ is the mixing parameter. In the calculations presented here, we use $\alpha=0.03$ and a residual tolerance $f_{\rm tol}=3\times10^{-7}$. The Dirichlet contact potentials are imposed exactly throughout the iteration. To facilitate convergence for large transverse sizes, we additionally use a simple continuation procedure \cite{AllgowerGeorg}, successively increasing the transverse size as $L_z=8~\mu{\rm m}\rightarrow20~\mu{\rm m}\rightarrow40~\mu{\rm m}$. The converged solution at each step is used as the initial condition for the next one.

\subsection{Convergence and charge neutrality}

Convergence is monitored through the electrostatic residual together with the corresponding variation of the charge density
\begin{equation}
    \delta\phi = \max_x \left|\widetilde{\phi}(x,0)-\phi(x,0)\right|, \qquad 
    \delta\sigma=\max_x\left|\sigma[\widetilde{\phi}](x)-\sigma[\phi](x)\right|.
\end{equation}
Any Gaussian smoothing applied to the final charge-density profiles is used only for graphical representation and does not enter the self-consistent calculation.

Finally, the Neumann boundaries at $z=\pm L_z/2$ carry no compensating surface charge since the normal electric field vanishes there. A nonzero net charge in the 2D layer is instead compensated by the induced charge on the lateral Dirichlet contacts. Defining 
\begin{equation}
    Q_{\rm sheet}=-e L_y \int_{-L_x/2}^{L_x/2}\sigma(x)\,dx,
\end{equation}    
global electrostatic consistency requires
\begin{equation}
    Q_{\rm sheet}+Q_{\rm contacts}\simeq0,
\end{equation}
up to the numerical convergence tolerance. This relation is explicitly checked for the converged solutions.

\label{app:capacitance_details}

\begin{figure}[h!]
    \centering
    \includegraphics[width=0.7\linewidth]{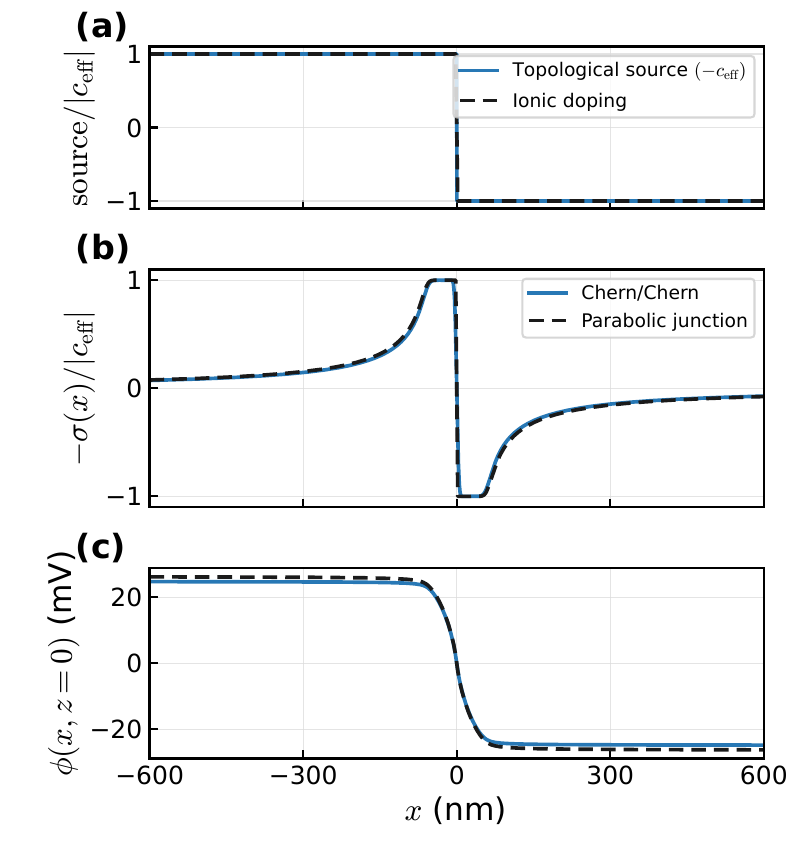}
    \caption{
Comparison between a topological Chern junction and a conventionally doped parabolic-band junction at $T=6.25\,\mathrm{K}$ and $B=1\,\mathrm{T}$. The topological model parameters are $A=150\,\mathrm{meV.nm}$, $B_m=-225\,\mathrm{meV.nm^2}$, $a=1\,\mathrm{nm}$, and $\Delta=50\,\mathrm{meV}$. The Dirac masses are $M_L=-25\,\mathrm{meV}$ and $M_R=-1775\,\mathrm{meV}$, corresponding respectively to low-energy valleys at $\Gamma$ and $\mathcal{M}$ and to Chern numbers $\mathcal{C}_L=-1$ and $\mathcal{C}_R=+1$. The mass varies across the interface over $\xi_M=10\,\mathrm{nm}$. The conventional parabolic-band junction has the same gap, $E_g=50\,\mathrm{meV}$, and a matched effective mass $m^\ast/m_e=0.0847$. In both cases, the background source has amplitude $|c_{\mathrm{eff}}|=2.42\times10^{10}\,\mathrm{cm}^{-2}$ and changes sign at the interface. (a) Normalized source profile $\mathrm{source}/|c_{\mathrm{eff}}|$. (b) Corresponding normalized self-consistent sheet charge density $\sigma(x)/|c_{\mathrm{eff}}|$. (c) Electrostatic potential $\phi(x,z=0)$ along the two-dimensional material. The two junctions exhibit nearly identical charge profiles, including a well-developed depletion plateau, and comparable built-in potentials, with $V_{\mathrm{bi}}\simeq49.7\,\mathrm{mV}$ for the Chern--Chern junction and $V_{\mathrm{bi}}\simeq52.6\,\mathrm{mV}$ for the parabolic junction.}
    \label{fig:topological_vs_doped}
\end{figure}

\section{Comparison: doped conventional junction vs. pristine Chern junction}
\label{app:comparison}

Figure~\ref{fig:topological_vs_doped} compares the self-consistently calculated charge density profile of a Chern junction with that of a junction made from doped conventional (non topological) parabolic band semiconductors. When the spatial profile of the dopant density, the energy gaps and the effective masses in the conventional junction match with the spectral asymmetry profile, the energy gaps and the effective masses of the Chern junction, the two systems yield the same spatial charge distribution.

\end{appendix}






\bibliography{SciPost_Biblio.bib}


\end{document}